\documentclass[preprint, twocolumn]{aastex701}

\newcommand{\Di}{\mbox{Di}}
\newcommand{\Ra}{\mbox{Ra}}
\newcommand{\Ek}{\mbox{E}}
\newcommand{\Pra}{\mbox{Pr}}

\newcommand{\Ro}{\mbox{Ro}}

\newcommand{\strat}{\mbox{A}}
\newcommand{\mpo}{\;\;.}
\newcommand{\mco}{\;\;,}
\newcommand{\dsc}{\partial s_c \big/ \partial r}

\received{XXXX, 2025}
\revised{XXXX, 2025}
\accepted{XXXX}
\submitjournal{ApJ}
\shorttitle{Atmospheric dynamics of Hot Jupiter }
\shortauthors{Dietrich et al.}
\graphicspath{{./}{figures/}}

\newcommand{\bs}[1]{\tilde{#1}}
\def\bm#1{\ensuremath{\mathchoice{\mbox{\boldmath$\displaystyle#1$}}
{\mbox{\boldmath$\textstyle#1$}}
{\mbox{\boldmath$\scriptstyle#1$}}
{\mbox{\boldmath$\scriptscriptstyle#1$}}}}
\usepackage{nicefrac}
\usepackage{journals}
\usepackage{mathtools}
\usepackage{placeins}

\newcommand{\WD}[1]{{\color{black} #1}}

\begin{document}


\title{Flow Regimes in Hot Jupiter Atmospheres: Insights from Anelastic Models}

\correspondingauthor{Wieland Dietrich}
\email{dietrichw@mps.mpg.de}
\author[0000-0002-5502-5040]{Wieland Dietrich}
\affiliation{Max Planck Institute for Solar System Research \\
Justus-von-Liebig-Weg 3 \\
37077 Goettingen, Germany}
\email[show]{dietrichw@mps.mpg.de}  

\author{Johannes Wicht}
\affiliation{Max Planck Institute for Solar System Research \\
Justus-von-Liebig-Weg 3 \\
37077 Goettingen, Germany}
\email{wicht@mps.mpg.de}

\begin{abstract}
Hot Jupiters are Jupiter-sized exoplanets with close-in orbits, characterized by extreme day-night temperature contrasts due to synchronous rotation. These planets offer unique observational opportunities through transit photometry, transmission spectroscopy, and infrared (IR) phase curve analysis, which reveal information about heat redistribution and atmospheric dynamics. 

Complementary to common generalized circulation models (GCMs), we introduce a more comprehensive approach using the anelastic fluid equations that fully capture the three-dimensional nature of the emerging non-linear flows. We identify various non-linear flow regimes and analyze the heat distribution when irradiation and thermal advection reach equilibrium.

Eastward zonal winds can reach velocities comparable to the planetary rotation (up to several kilometers per second), while slower radial flows, though less prominent, contribute significantly to heat advection and can cause both eastward and westward hotspot shifts. The efficiency of day-to-night heat redistribution and the positioning of brightness maxima are shown to depend strongly on pressure and the interplay of advective and radiative processes.

These findings improve our understanding of the diversity observed in the IR phase curves and suggest a non-magnetic mechanism for retrograde hotspot shifts. By extending the scope of traditional GCM models, our work demonstrates the usefulness  of anelastic models in capturing the complex, multidimensional dynamics of irradiated exoplanetary atmospheres.

\end{abstract}

\keywords{exoplanets --- atmospheric dynamics --- Gaseous planets --- numerical simulations}

\section{introduction}
Hot Jupiters (HJs), a class of exoplanets characterized by their close proximity to their host stars, offer a unique laboratory for studying extreme atmospheric dynamics. They are typically \WD{larger than} Jupiter but orbit their stars at distances much closer than Mercury to the Sun, often completing an orbit in just a few days. As a result, their atmospheres are subject to intense stellar irradiation, leading to extraordinary temperature gradients. Most HJs are tidally locked, meaning that one side of the planet is permanently facing the star while the other remains in permanent darkness. This leads to extreme day-night temperature differences, which in turn drive powerful winds and circulation cells that seek to redistribute heat. 
Understanding these atmospheric flows is crucial for interpreting observations of HJs since they influence both the thermal structure and the chemical composition of the planet’s atmosphere.

This study mainly aims to explore the key factors influencing atmospheric circulation in HJs with a particular focus on the morphology and stability of the emerging flows and the physical processes governing heat redistribution in these unique planetary environments. \WD{However, we study fundamental solutions for a wide range of parameters that more generally apply to the atmospheres of 
planets orbiting their host stars in synchronous rotation.}

\WD{While the dynamics is typically ruled by the interplay of the one-sided stellar irradiation and strong Coriolis forces, we find a wide range of solutions.}
This will help in understanding the diversity of the observed IR light curves \WD{of close-in exoplanets}. These
curves represent the vertically integrated three-dimensional temperature structure, 
which is shaped by advective flows seeking to equilibrate the one-sided stellar irradiation. 
A second observationally accessible quantity is the mean day-to-nightside difference quantifying the efficiency of redistributing heat from the day to the nightside. 

The impact of the stellar irradiation can be quantified by the 
radiative time scale $\tau_{rad}$ required to establish the thermal equilibrium where the local emission balances 
the stellar irradiation. This time scale increases strongly with pressure \citep{Showman2002, Iro2005}. At shallow depths or small pressures (above 10 mbar), the radiative time scale is short (less than $10^4$ seconds) and the stellar irradiation dominates. In deeper layers,  $\tau_{rad}$ increases beyond the advective time scale (between $10^5$ and $10^6$ seconds) and the thermal structure is shaped by advection. Hence it can be expected that thermal structure varies strongly with radius or pressure.

A strong impact of (horizontal) eastward advection has been theoretically predicted \citep{Showman2002} 
and seems to explain the observed eastward 
hotspot shifts in infrared phase curves taken by the Spitzer telescope, 
for example for HD 189733b \citep{Knutson2012}. 
The shift reflects the thermal advection by the dynamics in  all the layers relevant for the thermal emission. By now, hotspot shifts and day-to-nightside temperature differences 
have been inferred for several dozen HJs \citep{Parmentier2018, Bell2021}. The amplitudes of phase shifts range between 5 and 50 degrees longitude and can also be negative, implying a westward \WD{shift} \citep{Dang2018}.

Westward shifts have, for example, been attributed to the effect of Lorentz forces \citep{Rogers2014, Hindle2021,Boening2025}. 
\citet{Dietrich2022} showed that strong Lorentz forces are indeed a possibility for ultrahot Jupiters 
with equilibrium temperatures beyond about $1500\,$K. 
At such temperatures, the thermal ionization of alkali metals \citep{Kumar2021} becomes so efficient 
that the electrical conductivity reaches about $1\,$S/m, boosting magnetic effects. 
However, no simple relation of the hotspot offset or day-to-nightside difference on  planetary parameters like the semi-major axis, the orbital period, the planetary mass, the planetary radius, the equilibrium temperature, or the infrared wave-length has 
been found \citep{Parmentier2018, Bell2021}.  

Previous numerical studies of irradiation driven flows, pioneered by \citet{Showman2002}, are mostly based on  various flavors of general circulation models (GCM), each with its own set of simplifications and extensions (i.e. 2D shallow water \citep{Debras2020}, magnetic drag GCM \citep{Rauscher2013}, radiative hydrodynamics \citep{Dobbs-Dixon2013}, shock capturing $\beta$-plane approximation \citep{Fromang2016} or full MHD \citep{Rogers2014}). 
More details can also be found in the review papers by \citet{Showman2019} and \citet{Showman2020}. Most of these models applied the hydrostatic, primitive equations where the radial flow is a diagnostic variable rather than dynamically evolving under the influence of various forces. In some models, the irradiation driven dynamics are very time-dependent \citep{Cho2021, Rogers2014} and strongly affected by its fine-scale structure \citep{Skinner2025}.

\WD{More recently, it has been shown that thermal dissociation and recombination of hydrogen may significantly enhance lateral heat transport for ultra-hot Jupiters with dayside temperatures exceeding 2000\,K \citep{Mansfield2020}. However, we neglect this effect here. Another complexity that is ignored for simplicity, is the effect of clouds and condensation on heat transport and radiative transfer \citep{Parmentier2016, Heng2012}. The impact of both effects have already been studied with GCMs but not yet with anelastic models.}

Here we introduce an alternative approach where we model the irradiation driven dynamics of HJs on the basis of a hydrodynamic model that is more commonly used for convective atmospheres of giant planets \citep{Gastine2012, Dietrich2018, Wicht2020, Wulff2024} or dynamos in planetary cores \citep{Wicht2002, Dietrich2013}. Contrary to GCM models, the radial force balance is explicitly calculated considering all radial forces. This is possible by using the anelastic approximation, which ignores acoustic waves but captures a multitude of other atmospheric instability, such as thermal convection, Kelvin-Helmholtz instabilities, gravity waves or self-consistent magnetic effects. 
Here we focus on the hydrodynamic flows driven by the irradiation gradients in a stably stratified layer. We employ the numerical code MagIC, which has already been 
successfully applied to modeling stratified layers in various setups \citep{Wicht2002, Dietrich2018, Wulff2022}. The MAGIC-code is available at an online repository (\url{https://github.com/magic-sph/magic}). \\

\section{Properties of Hot Jupiters}

 \begin{figure}
     \centering
     \includegraphics[width=0.5\textwidth]{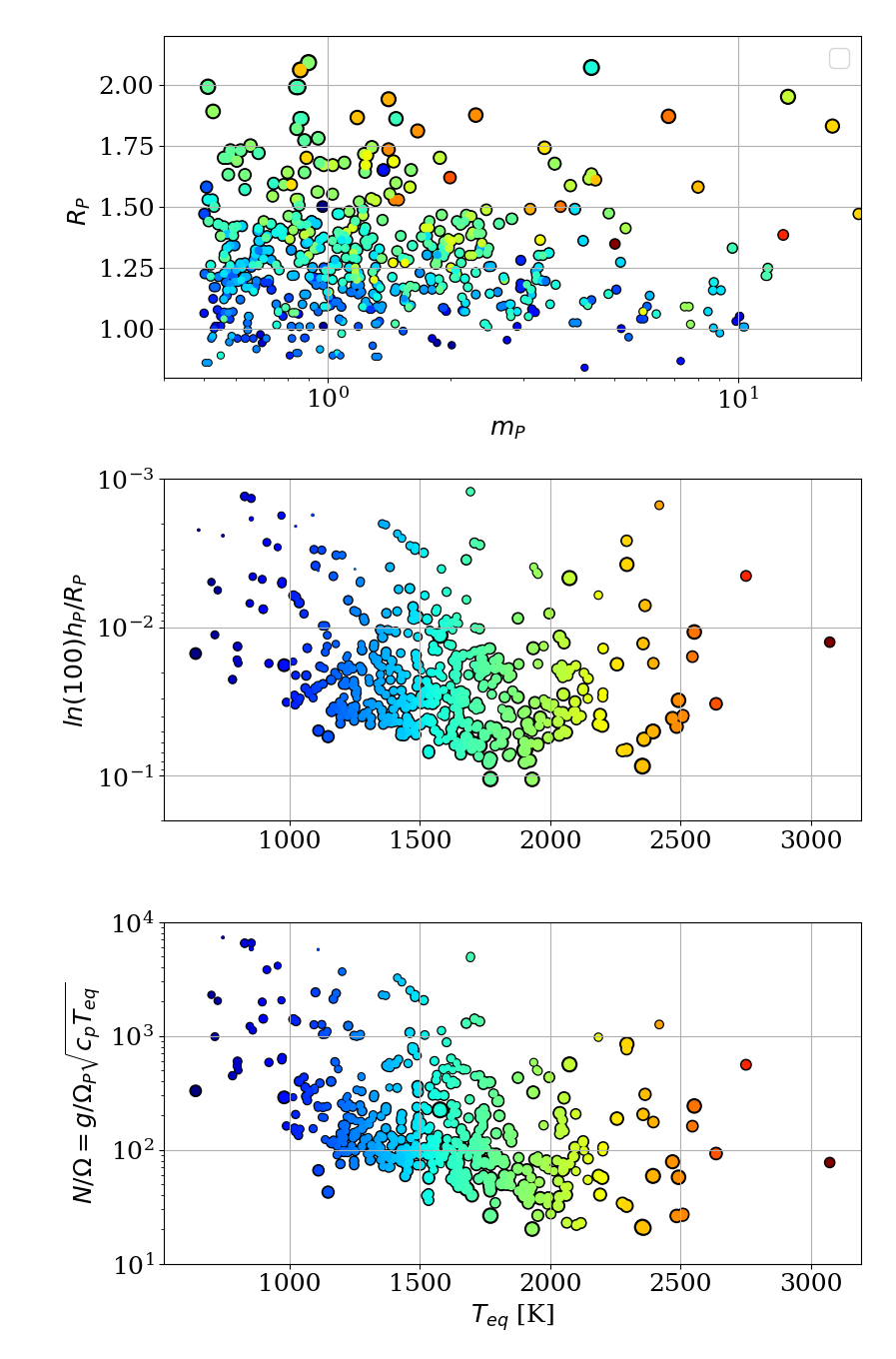}
     \caption{Statistics of characteristic HJ properties. Top panel: Mass-radius diagram.  
     Middle panel: Depth for a hundredfold pressure increase relative to planet radius. 
     Bottom panel: Ratio of the isothermal Brunt-V\"ais\"al\"a and the rotation frequency $N_T/\Omega$. 
     The size of the symbols is scaled with the planetary radius, while the color 
     indicates the equilibrium temperature.}
     \label{figHJintro}
\end{figure}

We define HJs as gas giants with a semi-major axis $\leq$ 0.1\,AU and more than
50\% of Jupiter's mass ($m_p \geq 0.5\,m_{jup}$).
Fig.~\ref{figHJintro}a) shows the mass-radius scatter plot for roughly 600 HJs in this category. 
The color of the symbol indicates the equilibrium temperature, the size of the symbol the planetary radius. Most HJs have a mass comparable or smaller than Jupiter, while the radius is up to two times larger. The mean density and the surface gravity are therefore typically much smaller than on Jupiter and seem to become even smaller 
with temperature (symbol color). 
On the permanent day-side, the radiative equilibrium temperature due to the intense stellar irradiation can reach several thousand Kelvin. Large atmospheric temperatures seem to correlate with low density, 'fluffy' planets. \WD{A recent review on the radius anomaly is provided by \citet{Thorngren2024}. Potentially contributing mechanisms are tidal dissipation \citep{Bodenheimer2001}, radiative transfer due to enhanced opacity \citep{Burrows2007}, Ohmic heating due to the dissipation of electric currents \citep{Batygin2010}, or heat advection \citep{Tremblin2017}. However, interpreting the observational data with respect to a single heating mechanism remains non-unique.}

 This seems to be somewhat more common for the hotter planets, where the irradiation is sufficient to thermally ionize atmospheric constituents and generate sizable electrical conductivities \citep{Kumar2021}, thereby promoting magnetic induction effects and possibly sufficient Ohmic heating to explain the excess radius \citep{Batygin2010}. However, such MHD effects may only be relevant when the atmospheric temperatures are large enough, i.e. exceeding 1500\,K \citep{Dietrich2022}. 
 This has been confirmed by simplified shallow water models 
 \citep{Hindle2021} and 3D simulations \citep{Rogers2014,Rogers2017,Rogers2017b}.
 Recently, \citet{Boening2025} used a setup similar to 
 this paper to explore magnetic effects in ultra HJs with 
 temperatures beyond $2000\,$K. They find a new magnetic instability 
 that is triggered in the presence of a background field representing 
 a deep planetary dynamo. This instability fundamentally suppresses the \WD{ generation of prograde equatorial jet} an\WD{d} can explain reduced o\WD{r} even westward hotspot shifts. 
 When the background field reaches at least $2.5\,$Gauss, the 
 imposed field sets off a subcritical dynamo that survives even
 when the background field is switched off. 
 
 Here we focus on the more numerous cases of HJs that can be considered hydrodynamic since their atmospheric temperature\WD{s} are below 1500\,K.
 Due to the strong irradiation parts of the outer atmosphere becomes isothermal 
\citep{Poser2019}. The radial pressure scale height, the characteristic scale height for the exponentially decreasing pressure, is then given by $h_P = k_B T / \mu g$, where $k_B$ is the Boltzmann constant, T temperature, $\mu$ the molar mass of the atmospheric constituents and $g$ the gravitational acceleration. 
 For relatively cold planets, like Earth or Jupiter, this length scale is at $8.5$ and $27\,\mathrm{km}$ and thus three orders of magnitude smaller than the respective planetary radii. However, for HJs the high atmospheric temperature and 
 smaller gravity can yield much larger pressure scale heights.  
 As two examples, the pressure scale height for HD 209458b and Kelt-9b are $720\,$km and $1140\,$km respectively. Fig.~\ref{figHJintro}b) shows the radius fraction over which 
 the pressure drops by two orders of magnitude, which typically amounts to 
 a few percent but can also exceed 10\% for some ultrahot Jupiters. 

\WD{An isothermal atmosphere is also stably stratified. It is characterized by the non-dimensional stratification parameter $N/\Omega$, i.e. the ratio
of the Brunt-Väisälä frequency $N$ and the rotation rate $\Omega$. 
It provides the ratio of negative buoyancy forces suppressing convection and the Coriolis force 
that promotes a stiffness of the dynamics and thus a tendency to penetrate 
stable layers.
As we will see below, it is crucially important for the dynamics of the system.}
Assuming  hydrostatic equilibrium, $N$ is determined by the entropy stratification of the atmosphere:
\begin{equation}
N^2= \frac{g}{c_p} \frac{d\bs{s}}{dr} = \frac{g}{T} \left( \frac{dT}{dr} -\frac{g}{c_p} \right)  \ ,
\end{equation}
where $s$ is the entropy per unit mass. 
However, since the temperature in the stratified part of the atmosphere is rather constant, $N$ can be approximated in the isothermal limit, \WD{ $N_T^2=g^2 / T c_p $}. Assuming $T=T_{eq}$ the relative
stratification parameter can be given by:
\begin{equation}
    \frac{N_T}{ \Omega}  = \frac{g}{\sqrt{ c_p T_{eq}}}  \frac{1}{\Omega} \;\;.
    \label{eqdefNome}
\end{equation}
\label{sec:Nome}
This measure is a particularly useful quantity since it characterizes the strength of the stratification and is accessible by observations. Assuming $c_p = 1.2 \cdot 10^4 \, \nicefrac{J}{kg\, K}$, as suggested by \citet{Poser2019}, yields typical $N_T/\Omega$ values between $30$ and 
$300$ as illustrated in Fig.\ref{figHJintro}c). \WD{This already indicates that  the radial flows are governed by the negative buoyancy due to the stable stratification dominates.}
Colder and thus more compact planets have a larger surface 
gravity and thus larger $N/\Omega$ values.

\section{Theory and numerical model}
\WD{The atmospheric flows in the outer part of the atmosphere of a HJ are driven by gradients in the stellar irradiation. We model the atmospheric dynamics as a rotating viscous fluid contained in a spherical shell with aspect ratio $a$ that is subjected to gravity forces and permanent dayside irradiation. 
Since magnetic effects may be important for the dynamics only for HJ at high temperatures \citep{Dietrich2022, Boening2025}, we ignore them here for simplicity. 

The general non-magnetic governing hydrodynamic equations are the conservation of mass, of momentum and of heat: 
\begin{eqnarray} 
\frac{\partial \rho}{\partial t} + \bm{\nabla} \cdot \left( \rho \bm{u} \right) &=& 0  \\
 \rho\left( \frac{\partial \bm{u} }{\partial t} + \bm{u} \cdot\bm{\nabla} \bm{u} \right) &=& -\bm{\nabla} p - 2 \rho \bm{\Omega} \times \bm{u} + \rho \bm{g} + \rho \bm{F}_\nu  \label{eq_NST}\\ 
 \rho T \left( \frac{\partial s}{\partial t} + \bm{u} \cdot\bm{\nabla} s \right)  &=&  \bm{\nabla} \cdot \left( \kappa \bm{\nabla} T \right) + Q_\nu + I_r \label{eq_THEQ} \ .
\end{eqnarray}
Here, $\bm{u}$ is the velocity field, $\bm{\Omega}$ the angular rotation vector, $\bm{g}$ the gravity vector and ${\bf F}_\nu$ the viscous force. In the third equation, $s$ is the specific entropy, $\kappa=k \rho c_p$ the thermal diffusivity, $k$ the thermal conductivity, $Q_\nu$ the viscous heating and $I_r$ is the Newtonian cooling function that models stellar irradiation. 

\WD{Eq.~\ref{eq_NST} is the full Navier-Stokes equation that includes the full Coriolis force and the viscous force ${\rho\bf F}_\nu$. The viscous heating term $Q_\nu$ in eqn.~\ref{eq_THEQ} 
quantifies conversion of kinetic energy into heat via viscous friction.
Both, viscous force and viscous heating, depend on the  stress tensor $\sigma_{ij}$: 
\begin{eqnarray}
 {F}_{\nu,i} &=& \frac{1}{\bar{\rho}} \partial_j \sigma_{ij} \, \text{,} \quad 
Q_\nu =\sigma_{ij}\frac{\partial u_i}{\partial x_j}\\ 
\sigma_{ij}&=&\bar{\rho}\left(\frac{\partial u_i}{\partial x_j}+\frac{\partial u_j}{\partial x_i}-\frac{2}{3}\delta_{ij}\nabla\cdot{\bf u}\right) \mpo
\end{eqnarray} }

\WD{The coupled set of partial differential equations can be simplified in various ways. 
The hydrostatic primitive equations (HPE) represent by far the most commonly applied set 
of simplification in the context of atmospheric general circulation models (GCMs).  
These involve (1) the assumption of a hydrostatic balance, (2) the assumption of a shallower atmosphere, 
and (3) a simplification of the Coriolis term. 
The first approximation means that radial flows are not actively driven by buoyancy forces. 
The other two approximations imply a simplified treatment of depth dependence and sphericity.
This approach has its limitations, in particular when it comes to deeper atmospheres 
and dynamics where radial flows play a more active role \citep{Mayne2019}.} 


Here, we follow an approach that may better resemble the deep, stratospheric circulation driven by irradiation gradients of a HJ. We solve the full set of equations, presented above, for disturbances around
a steady state background. All variables are hence decomposed into this background state, denoted by over-bars, and
dynamic disturbance, denoted by primes: 
\begin{equation}
x=\bs{x}(r) + x^\prime(r, \theta, \phi, t) \ .
\end{equation}

The background state is hydrostatic  ($\bm{u}=\partial_t=0$) , obeying the simplified 
eqs.~\ref{eq_NST} and \ref{eq_THEQ}:
\begin{eqnarray}
\bs{\rho} \bs{g} &=& \partial\bs{p} / \partial r \mco\label{eq:PB}\\
 0 &=& \bm{\nabla} \cdot \left( \kappa \bm{\nabla} \bs{T} \right) \mpo
\end{eqnarray}
It furthermore is assumed to depend only on radius and to be adiabatic ($ds=0$), 
which has the consequence that the background density depends only on pressure. 

We assume an ideal, perfect gas, which obeys the equation of state
\begin{equation}
p = (c_p-c_v )\rho T  \mco
\end{equation}
where $c_p$ and $c_v$ are the specific heat capacities.
Writing changes in density and temperature 
in terms of pressure and 
entropy yields  
\begin{eqnarray}
    \frac{d \rho}{\rho}  &=& - \frac{c_v}{c_p}\frac{d p}{p} - 
    \frac{d s}{c_p}  \mco \\ 
    \frac{d T}{T} &=& \frac{c_v-c_p}{c_p} \frac{d p}{p} + \frac{d s}{c_p}  \mpo
\end{eqnarray}
Using the hydrostatic equilibrium (eq.~\ref{eq:PB}) then 
provides the radial gradients of adiabatic background density and pressure: 
\begin{eqnarray}
    \frac{1}{\bs{\rho}} \frac{\partial\bs{\rho}}{\partial r}  &=& - \frac{1}{c_p - c_v}\frac{\bs{g}}{\bs{T}}  \ ,  \\ 
    \frac{\partial\bs{T}}{\partial r} &=& -\frac{\bs{g}}{c_p} \mpo
\end{eqnarray}

We also adopt the anelastic liquid approximation, discussed in detail by \citet{Braginsky1995} and \citet{Anufriev2005}, which neglects local temporal variations in density and 
thus any acoustic waves.  
The perturbations around the background state then obey the simplified evolution equations:
\begin{eqnarray} 
 \bm{\nabla} \cdot \left( \bs{\rho} \bm{u} \right) &=& 0 \nonumber \\
\frac{\partial \bm{u} }{\partial t} + \bm{u} \cdot\bm{\nabla} \bm{u}  &=& -\frac{1}{\bs{\rho}} \bm{\nabla} p^\prime - 2 \bm{\Omega} \times \bm{u} + \frac{\rho^\prime}{\bs{\rho}} \bs{\bm{g}} +  \bm{g}^\prime + \bm{F}_\nu  \nonumber \\ 
\frac{\partial s^\prime}{\partial t} + \bm{u} \cdot\bm{\nabla} s^\prime + u_r \frac{d\bs{s}}{dr}  &=& \frac{1}{ \bs{\rho} \bs{T} } \bm{\nabla} \cdot \left( \bs{k} \bm{\nabla} T^\prime \right) + Q_\nu + I_r \ , \label{heat2}
\end{eqnarray}
The perturbation of the gravity acceleration ${g}^\prime$ \WD{in the Navier-Stokes equation} can be eliminated by using the relation:
\begin{equation}
    -\frac{1}{\bs{\rho}} \bm{\nabla} p^\prime + \frac{\rho^\prime}{\bs{\rho}} \bs{\bm{g}} +  \bm{g}^\prime \simeq - \nabla \pi - \frac{s^\prime}{c_p} \bs{\bm{g}} \mpo
\end{equation}
where the reduced pressure $\pi$ is defined as
\begin{equation}
\pi = \frac{p^\prime}{\bs{\rho}}  + \Psi^\prime \ ,
\end{equation}
where $\Psi^\prime$ are the dynamic disturbances in the gravity potential, obeying $\bm{\nabla} \Psi^\prime = \bs{\bm{g}}^\prime$. More details on this derivation can be found in \citet{Braginsky1995} or \citet{Dietrich2021}.  
For simplicity we replace temperature fluctuations $T^\prime$ in the thermal equation (Eq.~\ref{heat2}) by entropy fluctuations $s^\prime$, which spares us from having to solve an additional equation for the temperature. 

The stellar irradiation $I_r$ is modeled with a Newtonian cooling term scaled by the pressure-dependent radiative time scale \citep{Showman2002, Iro2005}, 
\begin{equation}
    I_r = \frac{s^\prime-s_{eq}}{\tau_{\rm rad}(r)} \,
\end{equation}
where $s_{eq} \propto \cos \mu$ is the irradiation pattern, $\mu$ the stellar incident angle and 
$\tau_{\rm rad} = \tau_0 (\bs{p}/\bs{p}_o)^{\alpha}$ the radiative time scale. Here $\tau_0$ refers to the reference value at the outer boundary pressure $\bs{p}_o$ and $\alpha$ is the characteristic decay exponent. This reflects the pressure-dependent opacity increase with depth \citep{Iro2005, Tan2020}. We select to use $\alpha = 2$ to mimic a sharper decay that compensates the moderate pressure drop across the shell. \WD{This sharper increase of the radiative time scale is found for HD209458b at pressure of around 1 bar \citep{Iro2005}.
The pattern $s_{eq}$ represents the radiative equilibrium entropy that the system would assume without any 
atmospheric flows, where entropy production on the dayside and entropy loss on the nightside exactly 
balance.} 

\WD{Such a Newtonian cooling scheme offers a simple way of implementing radiation effects, 
which allows us to explore the fundamental dynamics for a wide range of parameters. 
More sophisticated approaches solve for radiation in two wavelength bands (double gray) 
or even multiple wavelength. A recent discussion can be found in \citet{Malsky2024}. }

\WD{The solutions of the equation system composed of Navier–Stokes and the heat equation are governed by the relevance of viscous and thermal diffusion, gravity, flow inertia, rotation rate, and irradiation strength, each characterized by a characteristic time scale. While some parameters can be specified realistically (e.g. the planetary rotation rate), others - such as molecular viscosity - cannot. However, it is the relative strength of these processes that determines the system’s behavior, which is naturally expressed by the ratio time scales, i.e. non-dimensional numbers. Rescaling the equations into non-dimensional form is therefore highly beneficial: it reduces the number of free parameters and highlights the key physical balances through dimensionless numbers (e.g. Rossby number $Ro=U/d\Omega$ ), enabling comparison across models and with astrophysical systems.}

We introduce the dimensionless formulation by choosing the shell thickness as length-scale $D=r_o-r_i$, the inverse rotation rate as time scale $t=\Omega^{-1}$, and an entropy scale of $D \,\vert \dsc \vert$, where $\dsc$  is the dimensional reference entropy gradient. This has the advantage that the 
numerous physical properties are collapsed into a few non-dimensional numbers that rule the dynamics. 

The non-dimensional form of the background state is then
\begin{equation}
    \frac{\partial \bs{T}}{\partial r} = 
    - \Di\;\bs{g}\mco
\end{equation}
\begin{equation}
    \frac{1}{\bs{\rho}}\frac{\partial\bs{\rho}}{\partial r}= 
    - \frac{\Di}{\Gamma}\,\frac{\bs{g}}{\bs{T}}\mpo
\end{equation}
introducing the Dissipation number
\begin{equation}
    \Di = \alpha g_o d / c_p\mco
\end{equation}
and the Grueneisen parameter, $\Gamma = (c_p-c_v) / c_v$. 

When dropping the primes of the disturbances 
the non-dimensional evolutions equations read: 
\begin{widetext}
\begin{eqnarray} 
\bm{\nabla} \cdot \left( \rho \bm{u} \right) &=& 0 \mco \\
 \frac{\partial \bm{u} }{\partial t} + \bm{u} \cdot\bm{\nabla} \bm{u}&=& -\bm{\nabla} \pi - \bm{\hat{e}}_z \times \bm{u} + \frac{\Ra\, \Ek^2}{\Pra} \frac{r_o^2}{r^2} \bm{\hat{e}}_r s  + \bm{F}_\nu \label{eqNST} \mco \\ 
 \frac{\partial s}{\partial t} + \bm{u} \cdot\bm{\nabla} s  + \strat u_r &=& \frac{\Ek}{\Pra} \frac{1}{\bs{\rho} \, \bs{T}} \bm{\nabla} \cdot \left( k \bm{\nabla} s\right) + \frac{\Ek}{\tau_0} \frac{s_{eq} - s}{ \bs{\tau}_{rad}}+ \frac{\Pra \, \Di}{\Ra\ \Ek} Q_\nu  \mpo
\label{eqheat}
\end{eqnarray}
\end{widetext}

The non-dimensional control parameters are the Rayleigh number, $\Ra$, the Prandtl number, $\Pra$, the Ekman number, $\Ek$, and the non-dimensional irradiation time scale $\tau_0$:
\begin{equation}
    \Ra = \frac{g_0 d^4}{c_p \nu \kappa} \big\vert \dsc  \big\vert \, , \, \Pra = \frac{\nu}{\kappa} \, , \, \Ek=\frac{\nu}{\Omega d^2} \, , \, 
  \tau_0 =  \tau_{rad}(0) \frac{\nu}{d^2} \ .   
\end{equation}
\WD{Each of these numbers quantifies the relative importance of the terms in the above equation 
and can be interpreted as ratios of different time scales. The Ekman number $\Ek$ is equivalent to the ratio of viscous diffusion and rotational time scales and essentially quantifying the relative importance of viscous effects. 
The Prandtl number $\Pra$ is the ratio of viscous to thermal diffusion. Since our setup is stably stratified and flows are driven by imposed irradiation gradients rather than convection, the Rayleigh number $\Ra$  reflects the efficiency of gravity in converting irradiation-induced entropy variations into radial 
motions. In the thermal equation, the stratification is controlled by the stratification parameter,
\begin{equation}
    A=(\partial \bs{s}\big/\partial r)\bigg/\left(\dsc\right) \ .
\end{equation}
}

The relative stratification given by $N/\Omega$ already 
introduced above can be then written in terms of the other non-dimensional parameters:
\begin{equation}
   \frac{N}{\Omega} = \Ek\;\sqrt{\frac{\Ra}{\Pra} \strat} \mpo
   \label{eqdefNOme}
\end{equation}

In the scaling used here, velocities are given in fractions of the rotational velocities and therefore have the values of so-called Rossby numbers, 
\begin{equation}
    \Ro=\frac{U}{D\Omega}\mco
\end{equation}
where $U$ is the velocity amplitude. It is the natural choice in this context, because it measures the ratio of inertial to Coriolis forces, i.e. the central balance that governs rotating astrophysical and geophysical flows.

The equations are solved by a modified version of the MagIC code \citep{Wicht2002, Gastine2012}. MagIC is a fully nonlinear, three-dimensional, pseudo-spectral code that solves for flow and magnetic field generation in a rapidly rotating spherical shell. To ensure mass conservation ($\nabla \cdot (\bs{\rho} \bm{u}) =0$), the 3D mass flux is written in terms of a poloidal potential $W$ and a toroidal potential $Z$:
\begin{equation}
    \bs{\rho} \bm{u} = \bm \nabla \times ( \bm\nabla \times W \hat{e}_r ) + \bm \nabla \times Z \hat{e}_r \ . 
    \label{eq:defpoltor}
\end{equation}
A toroidal mass flux has no radial contribution. This decomposition can be seen as the 3D generalization of the Helmholtz decomposition ($\bm{u} = \bm{\nabla} \chi + \bm{k}\times\bm{\nabla}\psi$) commonly used in atmospheric physics, where the flux is split between a divergent and rotational part \citep{Hammond2021,Sainsbury-Martinez2024}. Assuming $\bm{k}=\hat{e}_r$ yields $\psi=-Z$ and $\chi=\partial W /  \partial r$.  A toroidal mass flux is thus equivalent the rotational flux in the Helmholtz decomposition, while a poloidal mass flux is the 3D counterpart of the divergent flux. 

All scalar potentials are expanded in spherical surface harmonics $Y_\ell^m$ of spherical harmonic degree $\ell$ and azimuthal order $m$ via:
\begin{equation}
    Z (r,\theta,\phi) = \sum_{\ell = 0}^{\ell_{max}} \sum_{m = -\ell}^\ell Z_{\ell m} (r) Y_\ell^m (\theta, \phi) \ .
\end{equation}
The total poloidal and toroidal kinetic energies are then given by:
\begin{eqnarray}
    E_{pol} &=& \frac{1}{2} \sum_{\ell, m} \ell_1 \int_{r_i}^{r_o} \frac{1}{\bs{\rho}} \left( \frac{\ell_1 }{r^2} W_{\ell,m}^2 + \left(\frac{d W_{\ell,m}}{dr} \right)^2\right) dr \nonumber \\
    E_{tor} &=& \frac{1}{2} \sum_{\ell, m} \ell_1 \int_{r_i}^{r_o} \frac{1}{\bs{\rho}}  Z_{\ell,m}^2 dr \ ,
\end{eqnarray}
where $\ell_1 = \ell (\ell +1)$. 

This numerical technique has been successfully applied for modeling dynamos of solar systems planets and the global atmospheric hydro- and magnetohydrodynamics of giant planets \citep{Gastine2012, Wicht2020}. In particular, the MagIC code also been used for modeling  the dynamics in 
systems with partial stable stratification \citep{Dietrich2018, Gastine2021, Wulff2022}. 

The numerical models are setup by selecting the nondimensional numbers. The Ekman number of HJs is typically 
as small as $\Ek\approx 10^{-17}$ due to the small viscosity of hydrogen \citep{French2012,Boening2025}. 
In numerical simulations, much larger values have to be used in order 
to suppress small scale motions that cannot be resolved with
the available numerical resources. \WD{We therefore implicitly assume that the 
unresolved small-scale motions are dynamically not relevant for the larger scale
dynamics presented here.} 
We have explored Ekman numbers $3 \cdot 10^{-4}$,  $10^{-4}$, 
and $3 \cdot 10^{-5}$. 

The Rayleigh number $Ra$ and imposed stratification $\strat$ have 
to be adjusted to the chosen Ekman number in order to guarantee 
that relevant dynamical regimes \WD{ characterized by the relative stratification 
$N/\Omega$ are explored. }
We vary the Rayleigh number from  $\Ra=10^5$ to $\Ra=10^9$
and the stratification from $\strat=1$ to $\strat=10^3$.
This allows us to cover values from $N/\Omega = 10^{-1}$ 
to $N/\Omega=10^3$. 
The Prandtl number is $\Pra=0.1$ in all simulations, 
while the aspect ratio is fixed to $a=0.85$. 
As mentioned above, we chose $\tau_0 = \Ek/10$. The dissipation number, $Di$, is set to 3.0, the Grueneisen parameter $\Gamma = 0.5$ 
are kept constant. The  background pressure $\bs{p}$ varies by two orders of magnitude across the shell \citep{Boening2025}. 
The numerical resolution is $N_r = 73$ or $145$ in the radial direction and $N_\phi = 512$ or $1024$ in the azimuthal direction. 

Altogether we have run 124 models. Each has first been integrated until a steady or statistically steady state has been reached and has then been run for an additional significant fraction of a viscous diffusion time to explore the typical (time-averaged) characteristics. \WD{Utilizing MPI-parallelization over the radial levels, each run took several days of runtime.}
}
\section{Numerical results}

\subsection{Flow regimes}

\begin{figure*}
     \centering
     \includegraphics[width=\textwidth]{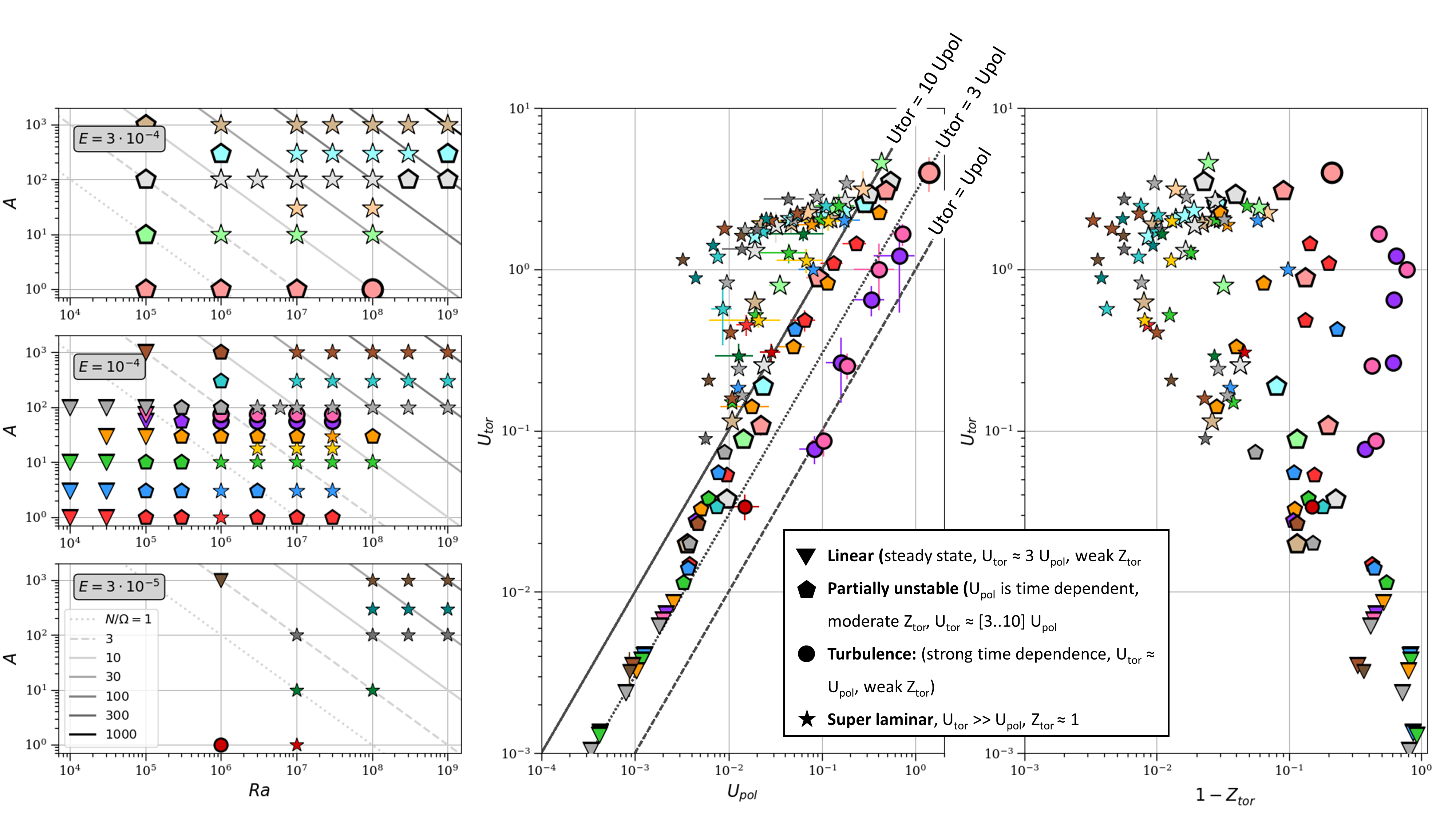}
     \caption{Overview of parameter regimes and types of solution.
     The three left panels show the location of the four solution types, indicated by different symbols, in the $Ra$/$A$ parameter space for three different Ekman numbers. 
     We classify the solutions into linear (triangles), partially unstable (pentagons), turbulent (circles) and super-laminar (stars).
     The symbol size is scaled with the Ekman number and the symbol color indicates the stratification $\strat$. 
     The center panel shows the characteristic non-dimensional poloidal $U_{pol}$ and toroidal flow amplitudes $U_{tor}$ \WD{scaled to provide Rossby numbers $\Ro=U/(D\Omega)$}. The right hand plot shows the zonality of the toroidal flow.}. 
     \label{fig:overview}
\end{figure*}

Fig.~\ref{fig:overview} provides an overview of all the performed simulations (124 in total). The left panels indicate the choice of Rayleigh number $Ra$ and stratification parameter $A$ for the three considered Ekman numbers $E$. The diagonal lines indicate the non-dimensional relative stratification $N/\Omega$ as given by eq.~\ref{eqdefNOme}. The large, middle plot in Fig.\ref{fig:overview} showcases the developed non-dimensional poloidal and toroidal flow amplitudes, $U_{pol}$ and $U_{tor}$, for all models. The diagonal lines indicate the ratio between the two. The error bars indicate the time variability ($\sigma$). The large plot on the right hand side illustrates the zonality, more precisely the deviation from the axisymmetry of the toroidal flow, $Z_{tor}$. Models with small $Z_{tor}$ feature very weak zonal flows, whereas in some of our models the toroidal flow is more than 99\% axisymmetric.

We classify the solutions based on the amplitude of toroidal and poloidal flow contributions and on the time dependence.  The flow amplitudes characterize the rms flow in the whole solution shell, 
\begin{equation}
    U_{tor} = \sqrt{\frac{\langle E_{tor}\rangle_t}{V} } E \quad \mbox{and} \quad  U_{pol} = \sqrt{\frac{\langle E_{pol} \rangle_t}{V} }  E \mco 
\end{equation}
where $\langle \cdot \rangle_t$ denotes the time average and $V=4\pi/3 \, (r_o^3-r_i^3)$ the shell volume. 
The temporal variation is quantified by the standard deviation of the poloidal flow: 
\begin{equation}
    \sigma_{pol}^2= \left\langle (E_{pol}(t) - \langle E_{pol} \rangle_t)^2  \right\rangle_t \mpo
\end{equation}
and similarly for the toroidal flow.

In classical convection, where heat enters from the bottom and escapes 
at the top, the Rayleigh number has to exceed a critical value to start motions. However, for the differential irradiation used here there is always motion. Flow speeds increase
generally with growing $\Ra$. The stable stratification primarily suppresses radial and thus poloidal flows, which are therefore typically smaller 
than their toroidal counterpart.

When the flows are slow, i.e.~for small $\Ra$ and/or large $\strat$, 
the solution remains linear and stationary. These linear solutions are indicated by triangles in Fig.~\ref{fig:overview} and are located in the lower left corner of the middle panel where both flow contributions are weak ($U_{tor} < 10^{-2}$).  The right panel shows that these solutions are also nearly purely nonaxisym\WD{m}etric (negligible $Z_{tor}$). Toroidal flows are typically three times faster than poloidal flows as indicated by dotted line in the middle panel.  
The spatial structure of the linear solutions depends on the relative strength of the Coriolis force and on the degree of stratification. The flow morphology is discussed in section \ref{sec:morph}.

When the flow amplitudes increase, non-linear terms become 
more important. This allows feeding energy into other modes, for example into the fast axisymmetric zonal flows, which circumvent the planet, or into various instabilities further described below.
For larger values of $\strat$, the
classical zonal-flow dominated solution emerges that has already been found in GCMs \citep{Showman2002}. Characteristic for these solutions is the clear dominance of zonal flows, indicated by a much larger 
toroidal than poloidal flow amplitude and by a high degree of zonality
$Z_{tor}>0.99$. 
In Fig.~\ref{fig:overview} we indicate respective solutions 
where $U_{\rm tor}\ge 100\,U_{\rm pol}$ by a star symbol. 
These solutions are super-laminar, having a very simple spatial structure and basically no time dependence. 

When increasing the Rayleigh number, $Ra$, the toroidal flow amplitude first grows but then saturates to a value somewhat larger than one (in terms of Rossby number), while $U_{\rm pol}$ keeps on growing. The
zonal-flow dominated super-laminar solutions there require 
a larger, \WD{b}ut not too large, $Ra$ and a high $\strat$ in order 
to suppress instabilities. 
At $Ro\approx1$ the Reynolds number is given by the inverse 
Ekman number. At $\Ek=10^{-4}$ we therefore have already flows with $Re=10^4$, 
a value where we expect turbulence or at least some kind of 
instability. However, the strong stable stratification at large $A$ keeps the solution laminar. 

Pentagons in Fig.~\ref{fig:overview} denote cases 
for $U_{\rm tor}< [3..10]\,U_{\rm pol}$ where the poloidal flow is rather
time dependent, typically given by $\sigma_{pol} > 0.01 \, U_{pol}$.
The time dependence is indicated by the \WD{vertical and horizontal bars} in the \WD{middle}  panel of Fig.~\ref{fig:overview}.
At low $\strat$ we find these solutions at the 
transition between linear solutions (triangles) and super-laminar solutions (stars). They also appear when increasing $Ra$ for 
a super-laminar case and thereby boosting the poloidal flow and 
approaching a Rossby number of order one. 
An example is the parameter combination $\Ek=3\cdot10^{-4}$,  $\strat=100$ and $\mathrm{Ra}=3\cdot10^8$. 
Characteristic are zonal flow contribution which are quite substantial with $Z_{tor} \approx 0.9$ but definitely lower than in the super-laminar regime. Also characteristic are small scale instabilities and waves, which typically emerge in the poloidal flow and explain the 
increase time dependence. We further discuss the possible 
instabilities below.

For $\Ek=10^{-4}$ and values of $\strat$ between 
$30$ and $100$ we also find particularly turbulent
solutions where toroidal and poloidal flow are of similar amplitude and very time-dependent, indicating another instability. \WD{In this distinct regime, the solutions have exceptionally high poloidal energy and are devoid of zonal jets, indicating a state of uncorrelated turbulence.} These solutions are indicated by the pink and purple circles in Fig.~\ref{instab2} and are discussed in more detail below. 
We did not find these solutions for $\Ek=3\cdot 10^{-5}$, possibly because of the stronger impact of rotation. 

\subsection{Flow morphology}
\label{sec:morph}
\begin{figure*}
     \centering
     \includegraphics[height=0.95\textheight]{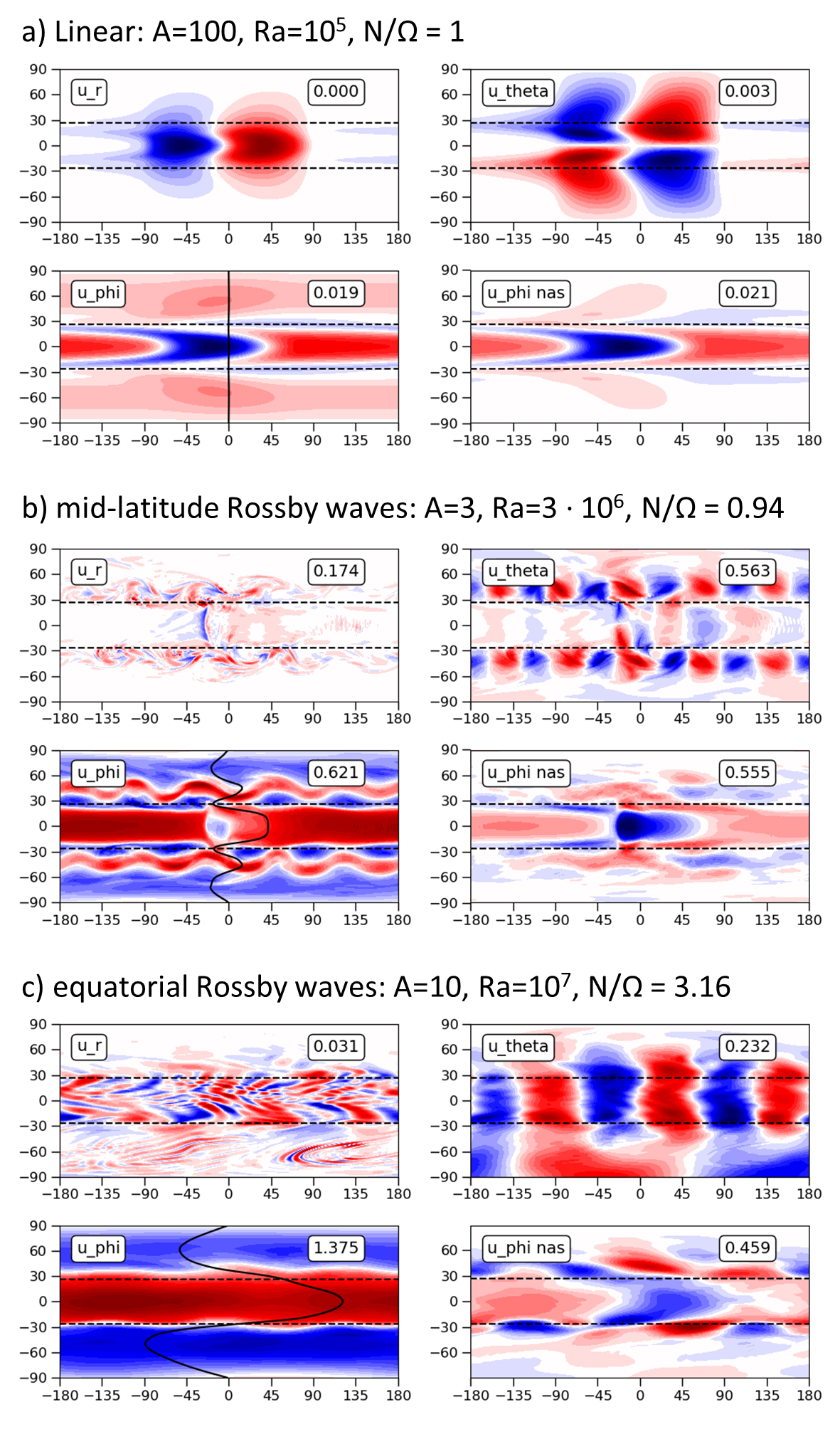}
     \caption{Flow solutions classified by spherical surface plots of $u_r$, $u_\theta$, $u_\phi$ and the nonaxisymmetric part of $u_\phi^\prime$ at $r/r_o=0.95$ }
     \label{instab1}
\end{figure*}

We will now discuss the structure and time-dependence of the different flow morphology types identified in our parameter scan. We concentrate on $\Ek=10^{-4}$ where we performed the most extensive 
study in terms of $Ra$ and $A$. 
Fig.~\ref{instab1} illustrates three different flow morphologies in
panels a), b) and c). 
For each, we show snapshots of all three flow components 
($u_r$, $u_\theta$ and $u_\phi$) and of the nonaxisymmetric zonal 
flows (lower right panels) on a spherical surface at a radius of $r/r_o=0.95$. 
Red colors refer to positive directions (radially outward, equatorward in the northern hemisphere and eastward), while blue color indicated negative flow directions (radially inward, poleward in the northern hemisphere and westward). Maximum flow amplitudes are given in the white boxes and the horizontal dashed lines mark the latitude of the tangent cylinder \WD{or TC} - the virtual cylinder that touches the inner boundary equator. \WD{In rapidly rotating systems, the flows are invariant along the rotation direction. Hence the TC marks the interface between {\it outside} where flows can extend vertically throughout both hemispheres. This is equivalent to the equatorial region up to latitudes where the TC intersects with the outer boundary. {\it Inside}, refers to latitudes closer to the poles, where the hemispheres are disconnected.}

Figure \ref{instab1}, a) shows the linear solution obtained by choosing small $Ra$. The radial flow (upper left panel) is set by the gradients of the irradiation pattern and thus inward west of the 
sub-stellar point (SSP) and outward east of it. The correlation of $u_r$ and $\partial_\phi s^\prime$ is expected from the linear theory due to the dominance of Coriolis forces \citep{Dietrich2016}. This is accompanied by one anti-cyclonic circulation cell in each hemisphere.
Both combine in the equatorial region to 
a strong westward (blue) flow through the SSP. 
The cells are mostly toroidal, representing 
equatorially symmetric circulations on a spherical shell.  \WD{As radial flows are rather weak, there is no meridional circulation. Moreover equatorward latitudinal flows are required east of the SSP in order to close the circulation cell.}
The zonal mean flow remains weak. Such linear solutions are shown as triangular  symbols in Fig.~\ref{fig:overview}.

Figure \ref{instab1}, b) shows a distinctly nonlinear case. The radial flow (upper left panel) develops a string of smaller-scale vortices at latitudes of $\pm35^\circ$ in addition to a persistent thin stripe of inward flow slightly west of the SSP and confined to the region outside of the tangent cylinder. 
Azimuthal flows (lower left panel) show a band of fast \WD{east}ward flows \WD{inside} of the 
tangent cylinder. This is flanked by two retrograde and one prograde 
jet in each hemisphere. The flanking jets have a wave pattern with
wave number $m=5$. The waves also clearly show in the latitudinal 
flows (upper right panel) and form a string of equatorially antisymmetric toroidal circulation cells mostly inside the 
tangent cylinder. 
The prominent wavy prograde jets, located between 30 and 60 deg latitude in each hemisphere, are reminiscent of the meandering jet streams in Earth's atmosphere. In addition to the strong prograde equatorial jet, a retrograde 
zonal flow patch remains anchored at the SSP and is 
reminiscent of the linear solution.

Analyzing the temporal behavior of these latitudinal features yields a westward drifting wave packet.  \WD{
We estimated a drift frequency of $\omega_{loc} = -0.039$ by tracking the local temporal evolution at a fixed point at $45^\circ$ latitude, sub-stellar longitude, and a radius of $r/r_o=0.95$.

Since the latitudinal features are not accompanied by 
related radial flows, the Coriolis force rather than gravity is likely the main restoring force, and Rossby waves are a likely candidate. Their characteristics (dispersion, propagation direction or frequency) are governed by azimuthal and longitudinal wave number, the background zonal flow profile and the rotation rate \citep{Ahlquist1982}.}

 Rossby waves obey the conservation of potential (total) vorticity and rely on spatial variation of the Coriolis force as a restoring force. Given in rotation time scales, the dispersion relation for Rossby waves in the $\beta$-plane approximation with a homogeneous background flow $U$ is:
\begin{equation}
\omega = U m - \frac{2 m}{\ell(\ell+1)} .
\end{equation}
This formulation does not take the stable stratification into account. 
Without background flow, the dispersion relation predicts 
a frequency of $-0.5$, clearly much faster than the wave in our
system. 
\WD{A prograde background flow ($U>0$) would reduce this value closer to the observed value. However, since the wave packet has a width of roughly 25 degrees in latitude, it experiences a complex background flow pattern that is partially nonaxisymmetric and even changes sign. The simple dispersion relation is thus
not applicable here and we expect that stratification and background flow may lead to significantly 
different eigenmodes.}

\WD{A dedicated study of possible eigensolutions may help here. \citet{Ahlquist1982} explores eigensolutions in a stably-stratified system geared to explain modes in Earth's atmosphere. For the azimuthal and longitudinal wavenumbers} $m=4$ and $\ell=4$ he reports a slow westward Rossby wave with $\omega=-0.05$, quite similar to the frequency identified here.
A calculation of the eigenvalues for the specific 
stratification and a realistic background flow in our setup, perhaps also for the background entropy, is required for a more 
convincing identification of the waves. 
However, this is beyond the scope of this paper. 

An increase in the relative stratification yields the solution
illustrated in Fig.~\ref{instab1}, c). The radial flows now spread along all longitudes yet remain mostly confined to equatorial latitudes \WD{outside} TC. 
The latitudinal confinement is likely due to the rotational constrain of the TC. 
The oscillations in the radial flow are likely gravity or inertial-gravity waves (IG waves). The typical horizontal wave number is $k_h =8$ or larger, and the drift is primarily retrograde. A typical frequency \WD{of the radial flow features} is obtained by averaging FFTs of the temporal evolution of the radial flow at a fixed point for all longitudes and peaks at $\omega_{loc} = - 0.15$. This is somewhat slower than expected for IG waves whose frequencies (in rotational time scales) \WD{range from unity to $N/\Omega$}.

Larger scale equatorially antisymmetric Rossby waves can again be identified in the latitudinal and the zonal flow. They now have of wave number $m=3$ and also fill the region \WD{outside} the tangent cylinder. \WD{The latitudinal flows actually cross the equator and connect both hemispheres.} The frequency of this wave with $m=3$ 
and a latitudinal wave number closest to $\ell=2$ is $\omega_{loc} = -0.07$. 
This is within the suggested range of frequencies for \WD{Rossby waves in Earth's atmosphere  \citep{Ahlquist1982}}. A second $m=1$ wave can be identified at higher southern latitudes. \WD{Similar inertial waves are found in the polar region of the sun \citep{Liang2025}.}

Inside of the TC, the nonaxisymmetric zonal flow is clearly part of the $m=3$ wave. Outside of the TC, however, the negative patch in the azimuthal flow is still fixed to the irradiation patch. 

\begin{figure*}
     \centering
     \includegraphics[height=0.65\textheight]{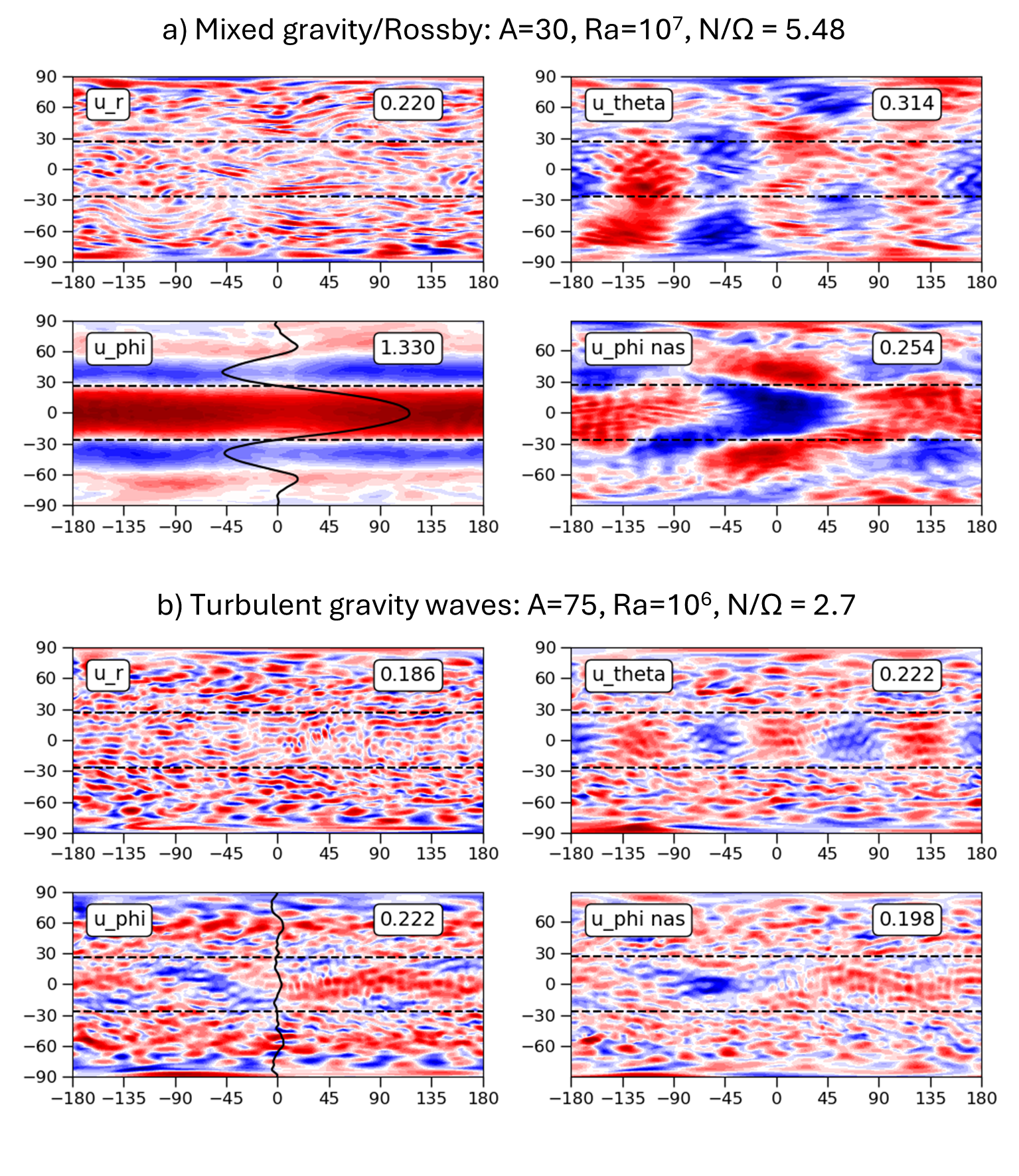}
    \caption{Like Fig.~\ref{instab1} but for 
    higher relative stratification.}  
     \label{instab2}
\end{figure*}

\begin{figure*}
     \centering
     \includegraphics[height=0.65\textheight]{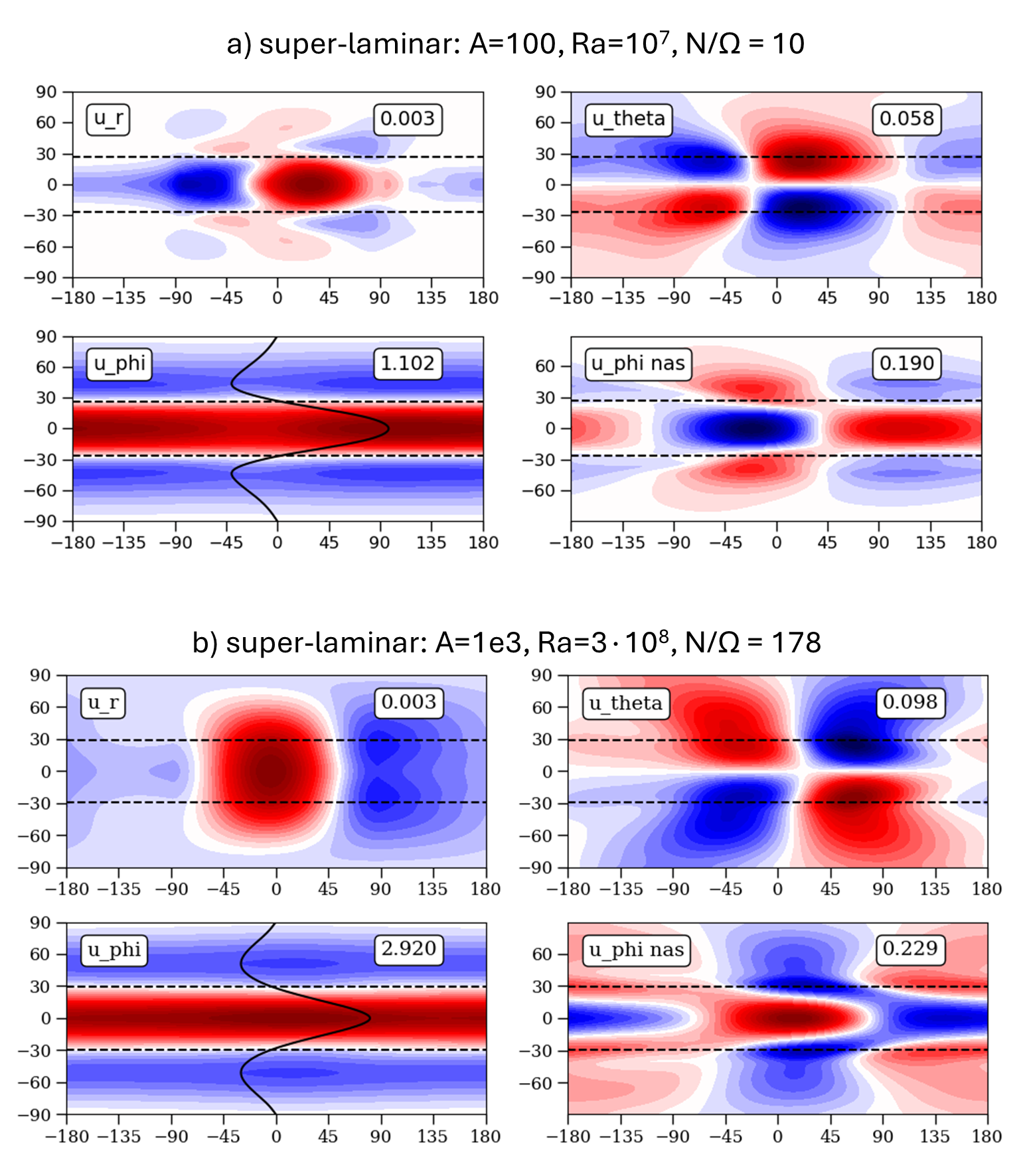}
    \caption{Like Fig.~\ref{instab1} but for the two configurations found in the super laminar regime.}  
     \label{instab3}
\end{figure*}

When increasing the stratification even further the flow becomes more small scale and shows more turbulent features. Fig.~\ref{instab2}, a) shows the solutions at $N/\Omega=5.48$. The radial flow is now dominated by small length-scale features that are distributed over the entire shell. The typical wave number is now $k_h \approx 30$. Again, we calculate the mean frequency with an FFT and find a relatively broad spectrum peaking at $\omega_{loc}\approx -0.5$. 

The {\it latitudinal flow} shows a superposition of the small-scale 
feature and the westward drifting Rossby wave with morphology of $m=3$
and (perhaps) $\ell=2$. The frequency for these slow, large scale Rossby wave is now about $\omega_{loc} = - 0.055$. Interestingly, the nonaxisymmetric {\it azimuthal} flow component in the bottom right panel  shows a mixture of similar features: the small-scale disturbance seen before in the latitudinal component, the large-scale slow drifting Rossby wave (in particular at higher latitudes) and the stationary patch \WD{of westward flow} anchored at the SSP. 
Despite the more complex nonaxisymmetric flow features, 
the correlation remains significant enough to drive 
strong zonal flows via Reynolds stresses. 
The lower left panel in Fig.~\ref{instab2}a) shows that additional
prograde (red) zonal flows have developed at high latitudes. 

When the relative stratification is increased to $N/\Omega=8.66$, 
however, the flows become so complex that the correlation is significantly reduced. The azimuthal flows are dominated by the small scale disturbance found in radial and latitudinal flows. The mean flow is weak in comparison, as is illustrated in Fig.~\ref{instab2}b). \WD{This means that the flow has lost its coherence and no Reynolds stresses can form.}
Such cases are depicted by circles in Fig.~\ref{fig:overview}; poloidal and toroidal energy are of similar magnitude and the zonality of the toroidal energy nearly vanishes.

In the FFT of the radial flow, the local mean frequency sharply peaks at $\omega_{loc}=0.47$, yet the propagation direction is less unidirectional. 
In the latitudinal direction, the large scale, equatorial symmetric Rossby wave remains visible in the equatorial region and features an $m=3$-mode with similar retrograde drift as identified before. 

The distinct scenario that we called super-laminar unfolds at even
larger values of $N/\Omega\ge10$, which are characteristic for HJs. 
The zonal (toroidal) flow amplitudes then 
dominate their nonaxisymmetric counterparts by two orders of 
magnitude (stars in Fig.~\ref{fig:overview}). 
Because of the strong stable stratification, the flows remain laminar 
despite Rayleigh numbers being large and the Reynolds numbers exceeding $10^4$.
Fig.~\ref{instab3}, a) illustrates a solution at $N/\Omega=10$ where 
no instability or wave is found and the 
solution is very simple and steady.
Radial outflows now concentrate 
at longitudes at and westward of the substellar point while
downflows can be found to the east of the outflows. The radial flows are a factor of 50 smaller than in the case discussed before. Since the strong stratification suppresses all 
time dependent small scale instabilities and reduces the radial flow amplitude, which 
is now by about a factor $50$ smaller the in the cases discussed before.
The azimuthal flows now show a very strong prograde jet at the equator that is flanked by two mid-latitude retrograde jets.

The nonaxisymmetric flows are organized in two large-scale cells, one in the northern and the other in the southern hemisphere. The center of these cells sits west of the 
SSP with equatorward flows around and east of the SSP, westward flows around the equator, 
and poleward flows towards the west of the SSP.
The cells are dominated by horizontal flows, but nevertheless strongly 
correlate with the much weaker radial flows. 
Equatorward flows strongly correlate with outflows, while poleward flows with downflows. 

\WD{Fig.~\ref{instab3}, b) shows another super-laminar flow scenario which 
preferentially emerges at rather large values of $N/\Omega$, here $178$.
While the zonal winds are very similar to the ones found in scenario a), 
radial outflows now clearly center around the SSP while the downflows can be found towards the 
east of the outflows. Once more, we find one mainly horizontal circulation cell in 
each hemisphere. However, equatorward flows are now  mostly located westward of the SSP and
mostly correlate with outflows. Poleward flows, on the other hand  can be found eastward of the SSP and mostly correlate with downflows. The circulation cells are closed by 
eastward flows in the equatorial region around the SSP.
This means that the northern cells turn clockwise in scenario a) but anticlockwise in scenario b). 
}

\begin{figure*}
     \centering
     \includegraphics[width=\textwidth]{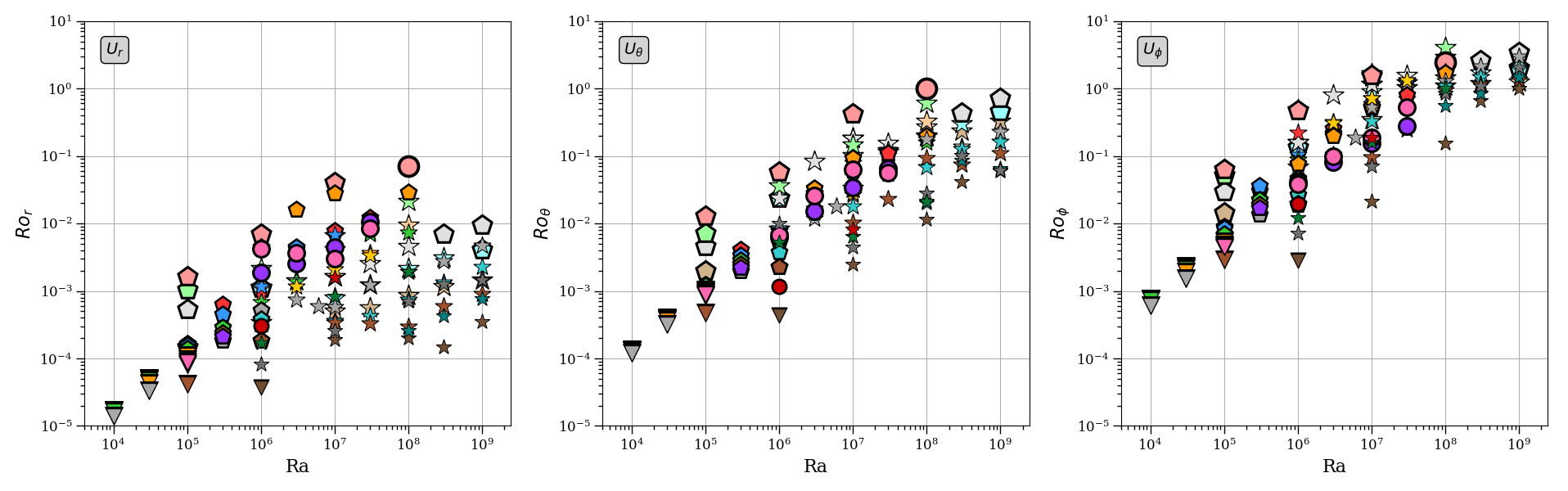}
     \caption{The rms flow amplitudes for all models as function of $\Ra$ in units of the Rossby numbers. Color and symbols are as in Fig.~\ref{fig:overview}. 
    }
     \label{fig_rms}
\end{figure*}

\WD{\subsection{Flow amplitudes}}
\WD{
The flow components are observationally more accessible than the poloidal and toroidal potential. 
Fig.~\ref{fig_rms} shows global rms averages of all three 
flow components in units of Rossby numbers, i.e. $Ro_r = U_r / \Omega R$, 
as a function of the Rayleigh number $Ra$ and the stratification parameter $A$. 
Generally, all components increase with Rayleigh number; almost linearly at small \Ra. The flows tend to become non-linear at high Rayleigh numbers, leading to some kind
of saturation. As a direct consequence of the stratification, the radial flows are generally smaller than the latitudinal or the azimuthal flows. 
The azimuthal flows (right panel) are mostly dominated by 
zonal jets, which are indirectly fed via the Reynolds stress mechanism described below. 

For the super-laminar cases (star symbols), which best resemble HJs in terms of $N/\Omega$, 
the radial, latitudinal and azimuthal flow amplitudes become independent of $Ra$ and saturate at values of $Ro_r\approx10^{-3}$, $Ro_\theta\approx0.1$ and $Ro_\phi\approx1$ 
respectively. Note that $U_\phi$ also becomes virtually independent of the stratification (colors). However, larger stratifications directly 
suppress radial flows and thus yield lower respective rms amplitudes. 

The Rossby numbers can be rescaled to physical velocities of a given planet by multiplying with the planetary rotation rate $\Omega_P$ and the planetary radius $R_P$. For typical values $\Omega_P$ and $R_P$, the azimuthal flow speeds amounts to km/s, the latitudinal flows to hundreds of m/s and the radial flows to a few m/s. This is in line with observations for azimuthal and radial flows \citep{Snellen2010, Komacek2019}. 

 }

\subsection{Driving zonal winds}

\label{sec:zonalflows}
\begin{figure*}
     \centering
     \includegraphics[width=\textwidth]{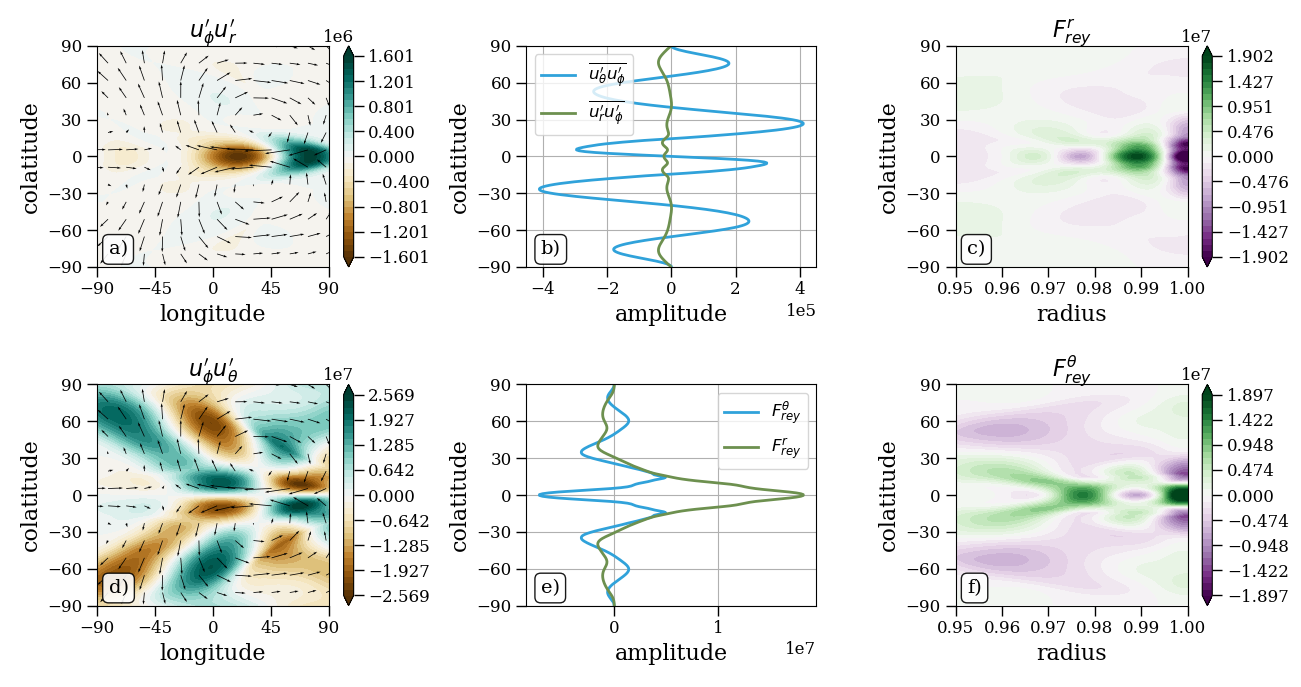}
     \caption{Spherical surface plots at $r/r_o =0.99$. Shown are the products of $u^\prime_r$ and $u^\prime_\phi$ (a) as well as $u^\prime_\theta$ and $u^\prime_\phi$ (d), the zonal mean correlation (b), 
     and the associated acceleration (e). Panels c) and f) show both acceleration contributions in 
     the meridional plane. Arrows illustrate the nonaxisymmetric circulation cells.}
     \label{figR1e8}
\end{figure*}

Investigating the acceleration and saturation of the zonal jets requires a more detailed analysis of the involved flows and forces.
The Reynolds stresses that drive the zonal winds arise from a
statistical correlations of the nonaxisymmetric flow components. 
Such a correlation can be achieved when the primary circulation cells are deformed by a prograde jet \citep{Showman2010}. \WD{In the context of Hot Jupiter in the super-laminar regime this has been often referred to as a standing wave.} A stronger deformation yields stronger Reynolds stresses and
thus faster jets. This leads to a run-away growth, which is 
ultimately balanced by viscous drag \WD{or other forces.}
 
Nonaxisymmetric flows $\bm{u}^\prime$ accelerate zonal winds, $\overline{u}_\phi$, 
if the zonal average of the azimuthal non-linear advection,
\begin{equation}
    \bm{F}_{rey}=\hat{\bm{e}}_\phi\cdot(\overline{\bm{u}^\prime \cdot \nabla u^\prime})
\end{equation}
does not vanish. A statistically persistent correlation of the non-linear flow components 
yields a weak but persistent acceleration that can ultimately results in fast zonal winds.
We can split the force into two components the depend either on the non-linear radial or the non-linear latitudinal flows:
\begin{eqnarray}
 \bm{F}_{rey} &=& \bm{F}^r_{rey} +\bm{F}^\theta_{rey}   \\
&=& - \frac{\sin \theta}{r^2} \partial_r \left( \bs{\rho} r^3 \overline{u_r^\prime u_\phi^\prime} \right) - \frac{\bs{\rho}}{\sin \theta}\partial_\theta \left( \sin \theta^2 \overline{u_\theta^\prime u_\phi^\prime} \right)\label{eqRey} \mpo
\end{eqnarray} 
Since this has the form of a gradient, the zonal mean of the product of the non-linear flows is often referred to as Reynolds stress.

Fig.~\ref{figR1e8} illustrates the different terms in eqn.~\ref{eqRey}
for a case at $Ra=10^8$, $A=100$ and $E=10^{-4}$ that 
is very similar to the super-laminar solution illustrated in Fig.~\ref{instab2}. 
The primary circulation cells, indicated by arrows in Fig.~\ref{figR1e8}, yield flow correlations 
that are mostly concentrated around the equator for $u_r^\prime u_\phi^\prime$ (panel a) but are 
complex and distributed for $u_\theta^\prime u_\phi^\prime$ (d). The azimuthal means are 
small for the first but large and highly latitude dependent for the second contribution (b). 
Because of the strong radial dependence of the flows the largest acceleration 
is nevertheless associated to the product $u_r^\prime u_\phi^\prime$. 
Except around the equator, the second acceleration due to the term $u_\theta^\prime u_\phi^\prime$ is generally similar in amplitude. Thus, both the depth dependence and the latitudinal deformation of the circulation 
cells into a chevron-like pattern may contribute to driving zonal flows. 
Panels c) and f) finally illustrates how both accelerations contribute at different
depth's and latitude, providing a rather complex picture.

\begin{figure*}
     \centering
     \includegraphics[width=0.95\textwidth]{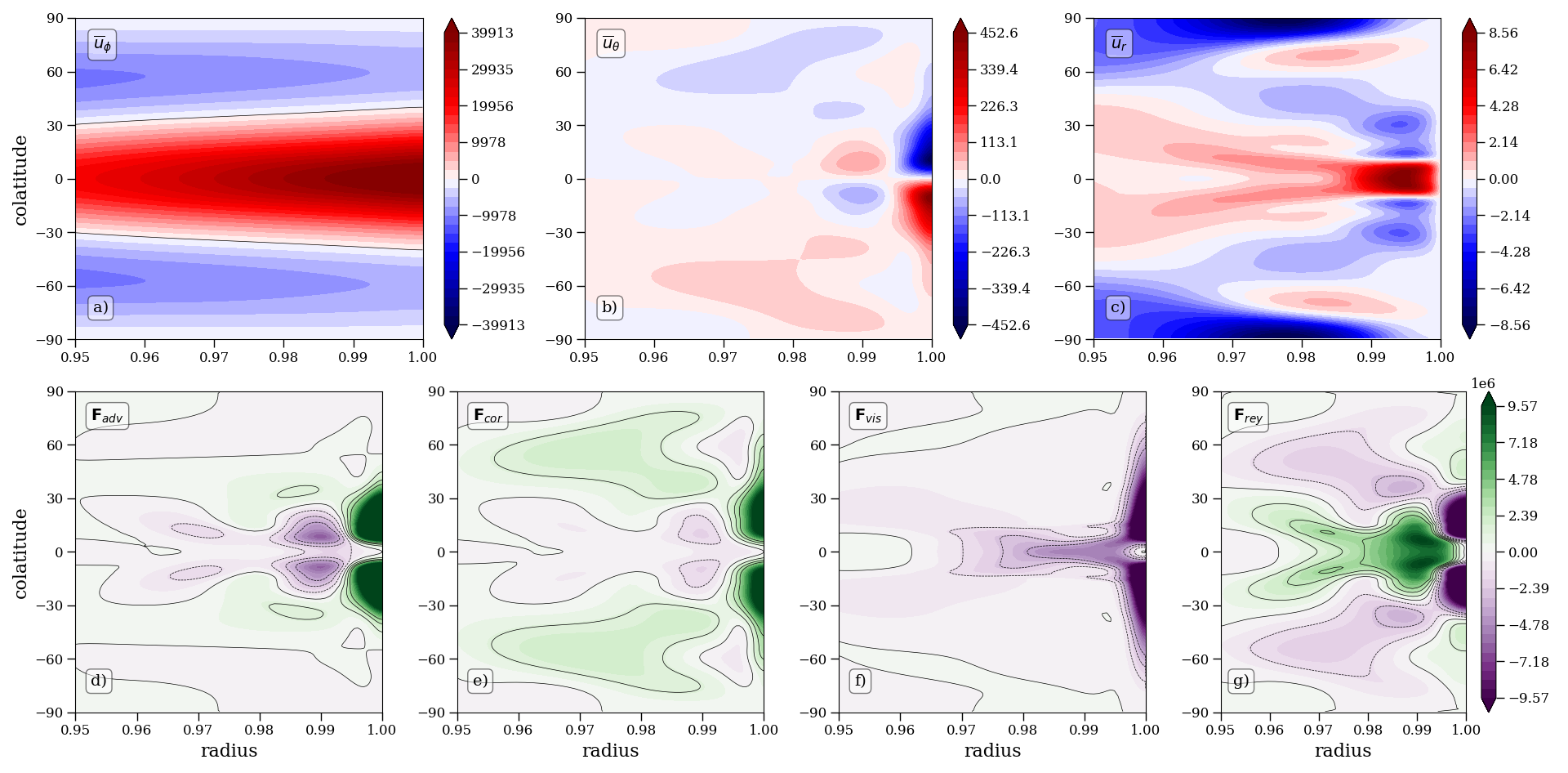}
     \caption{To\WD{p} row shows the zonally averaged azimuthal (a), latitudinal (b) and radial (c) flow. The bottom row highlights the acting forces, i.e. advective force (d), Coriolis (e), viscous force (f), and  Reynolds stress divergence (g). Parameters: $Ra=10^8$, $A=100$, $E=10^{-4}$. Green (purple) colors indicate acceleration (braking) of the local zonal flow.}
     \label{figAngMom}
\end{figure*}

Fig.~\ref{figAngMom} shows the axisymmetric contributions in all three
flow directions for the case already depicted in Fig.~\ref{figR1e8}. 
The prograde azimuthal jet around the equator becomes 
narrower with depth and generally slows down (panel a). 
The retrograde flanking
jets, on the other hand, become wider and faster with depth. 
Radial and latitudinal axisymmetric flows (panels b) and c)
yield one primary low-to-mid-latitude meridional circulation
cell in each hemisphere with 
upflows around the equator and 
downflows between  $10^\circ$ and $40^\circ$ 
northern and southern latitudes. 
Secondary cells, churning in the opposite direction, flank the primary cells. 
The meridional circulation is two orders of magnitude slower than the zonal winds. 


The balance between Reynolds stress acceleration and other forces determines the 
actual zonal flow configuration. 
Panels d) to g) in Fig.~\ref{figAngMom} shows all time-averaged 
axisymmetric azimuthal forces. 
The respective force balance reads 
\begin{equation}
0 = \bm{F}_{rey} + \bm{F}_{vis} - \bm{F}_{cor} - \bm{F}_{adv}    \mco
\label{eqconsL}
\end{equation}
where $\bm{F}_{adv}$ is the so called advective force, 
\begin{equation}
\bm{F}_{adv} = \bm{u}_m \cdot \bm{\nabla} \left( \frac{\overline{u}_\phi }{s} \right) 
\end{equation}
with $\bm{u}_m$  being the axisymmetric meridional circulation. The advective force describes the redistribution of zonal flow  by the meridional circulation cells, 
$\bm{F}_{rey}$ is the Reynolds acceleration already discussed above, and 
$\bm{F}_{cor}$ is the Coriolis acceleration and $\bm{F}_{vis}$ the viscous acceleration. 
Green colors show a local 
acceleration of the present zonal flow, while 
purple colors show local braking.  

There are two distinct regions in the force balance. At depths 
between $r=0.98\,r_o$ and $r=0.995\,r_o$, the prograde equatorial zonal wind is 
mostly driven (green color) by the Reynolds stress (panel (g) in Fig.~\ref{figAngMom}), while viscosity (panel f) and advective force (d) provide braking (purple). 
At very shallow depths, the advective force 
(panel d) and the Coriolis acceleration (panel e) take over powering the jet, whereas Reynolds acceleration (g) and viscosity brake (f). 
The advective force redistributes angular moment from the deeper region, where it is produced 
by the Reynolds acceleration, to the shallower region via the primary meridional circulation cells discussed  above. 

The location of the Reynolds acceleration (green color in panel g) shows that the contribution
$\bm{F}^r_{rey}$ (see panel c) in Fig.~\ref{figR1e8}), which is caused by the depth dependence of the flow, is most important 
for explaining the prograde jet in the equatorial region. 

 \begin{figure}
     \centering
     \includegraphics[width=0.95\columnwidth]{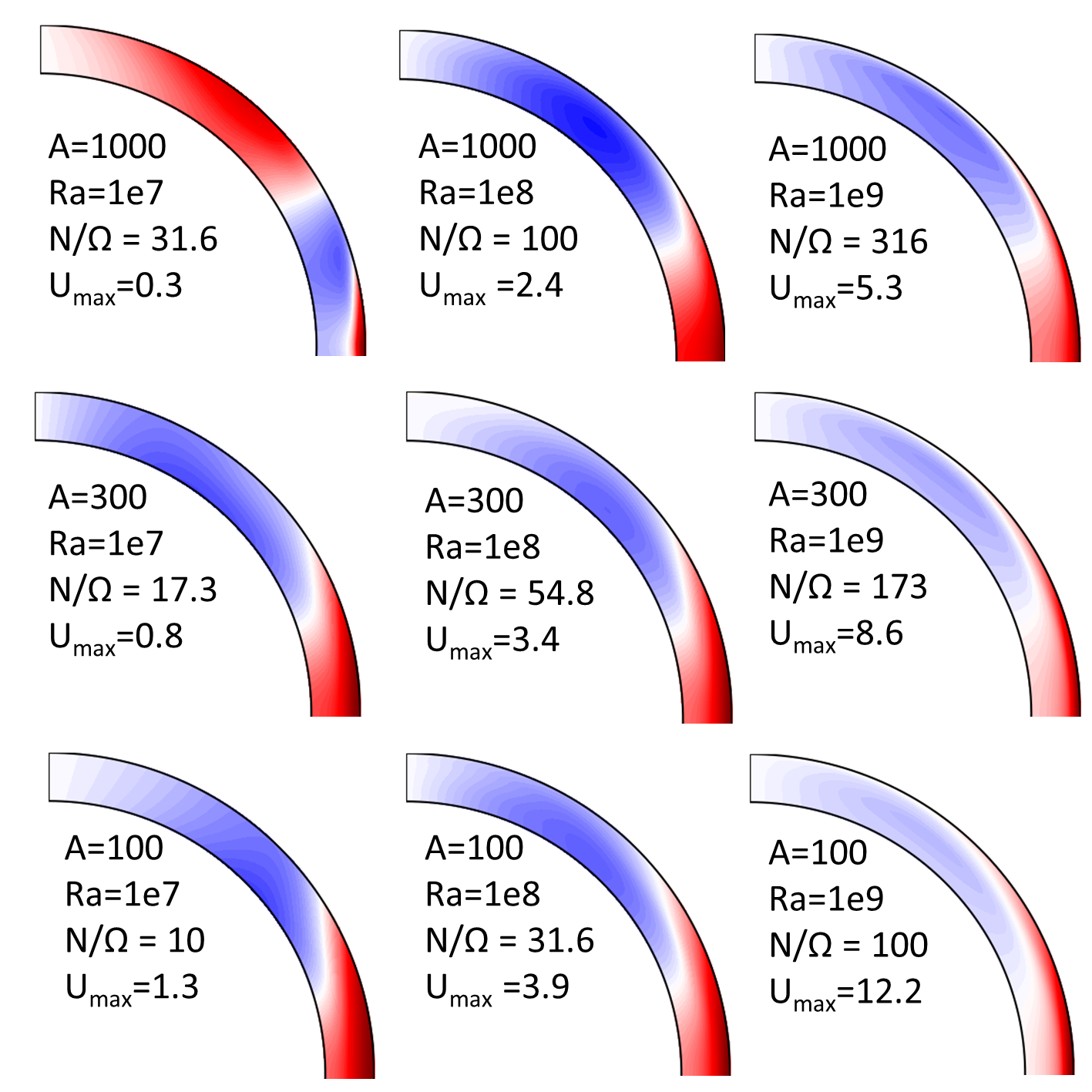}
     \caption{Various zonal flow patterns for models from the super-laminar regime at $\Ek=10^{-4}$. \WD{Plotted is the northern hemisphere of the meridional plane. Red colors indicate eastward, blue westward zonal flow.}}
     \label{figZonalFlow}
\end{figure}
\WD{The normalized stratification,  $N/\Omega$, controls the relative strength of stratification and Coriolis force. Rapidly rotating flows are dominated by the Coriolis force and hence tend to be invariant along the rotation axis, a phenomenon referred to as geostrophy.} 

Figure~\ref{figZonalFlow} illustrates various zonal flow patterns for models with realistic $N/\Omega$ values representative of HJs, corresponding to values of the stratification parameter between  $\strat=100$ and $\strat=1000$ and Rayleigh number between $\Ra=10^7$ and $\Ra=10^9$ at an Ekman number of $\Ek=10^{-4}$. 
The top-left model, characterized by $\strat=1000$ and $\Ra=10^7$, has not developed a prograde zonal jet near the equator. The strong stratification suppresses the formation
of jets and the solution remains in the 
linear regime despite the fact that the Rayleigh number 
is relatively high. 

In all other models illustrated in Fig.~\ref{figZonalFlow}, the typical intense mean zonal shear flow with a prograde flow at the equator and the retrograde higher latitude jets has formed. Models with smaller $\strat$ and higher Rayleigh numbers, $\Ra$, tend to develop stronger jets. As expected, the zonal jets become less geostrophic, i.e~vary more in the direction of the rotation axis, for larger values of $N/\Omega$, which increases with both $\strat$ and $\Ra$. 
The rotation-promoted geostrophy decreases, while 
the spherical geometry due to the radius dependent stratification 
becomes increasingly  obvious. 
While the equatorial jet is strongly geostrophic 
at $\strat=100$ and $\Ra=10^7$ with $N/\Omega=10$
(lower left panel in Figure~\ref{figZonalFlow}), 
i.e.~is nearly invariant in 
the direction of the rotation axis, 
it assumes a more spherical shape at 
$\strat=100$ and $\Ra=10^9$ with $N/\Omega=100$ (lower right panel). 
For the latter parameter combination, 
the jet is also faster and wider at the surface.
The latter is a consequence of the 
meridional 
advection of zonal flows towards higher latitudes
mentioned above.


\subsection{Advection of entropy}
\WD{The numerical solutions show complex, sometimes very time-dependent flows that can 
strongly modify the simple irradiated entropy pattern due to advective and dissipative 
effects. In this section we focus on the physical principles that shape the entropy pattern 
and are responsible for hotspot offset and 
day-to-nightside temperature equilibration.
We analyze the time-averaged entropy anomaly on
spherical surfaces, concentrating on models at $\Ek=10^{-4}$ that are in the zonal-flow dominated super-laminar 
regime (stars in Fig.~\ref{fig:overview}, middle panel in left column). 
These cases seem to best 
represent the conditions of HJs where phase curve properties are available \citep{Bell2021}.
Interestingly, we have found both eastward and westward hotspot offsets depending on the depth and 
on the non-dimensional parameters. 

In order to understand the dynamics that determines the entropy pattern 
we analyze the stationary entropy evolution (eqn.~\ref{eqheat}).
Newtonian cooling and flow advection clearly dominate, while 
viscous heating and thermal diffusion can be neglected in comparison.  
The equation thus reduces to 
\begin{equation}
u_r \strat\ + u_r \partial_r s + \frac{u_\theta}{r}  \partial_\theta s + \frac{u_\phi}{r\, \sin \theta}  \partial_\phi s \approx \frac{s_{eq}-s}{\tau_0 p^2} \mpo
\end{equation}
This can be further simplified by neglecting
the second term, the advection along the local radial
entropy gradients, in comparison to the 
first term, the advection along the 
imposed stratification gradient $\strat$. 
Though radial flows are slow, the latter term
remains important because of the large 
$\strat$ value. 
Since latitudinal flows are much slower than azimuthal flows, we 
can also neglect advection by the former flow component.
The entropy equation then simplifies to
\begin{equation}
\frac{s_{eq}-s}{\tau_0 p^2} - u_r \strat\ - \frac{u_\phi}{r\, \sin \theta}  \partial_\phi s \approx 0 \mpo
\end{equation}

Fig.~\ref{figadvection} illustrates the balance in the entropy equation for 
an eastward (left column) and a westward (right column) hotspot shift 
(see second panels from the top). 
The top panels show the depth dependence of the rms contributions of the terms in the entropy equation. 
In both cases, the advection by latitudinal flows is clearly subdominant.
Close to the top, Newtonian cooling always dominates and the 
entropy pattern is close to the equilibrium pattern. 
The entropy increasingly deviates from the radiation equilibrium with depth 
because of the decreasing Newtonian cooling and the growing impact of the dynamics. 

For the further analysis, we derive a latitudinally averaged entropy fluctuation 
\begin{equation}
    s^\star(r, \phi) = \int_0^\pi \left(s^\prime(r,\theta,\phi) - \tilde{s}(r) \right) \sin \theta d \theta \ ,
\end{equation}
where $\tilde{s}$ is the spherically symmetric entropy anomaly:
\begin{equation}
\tilde{s}(r) = \int_0^\pi \int_0^{2\pi} s^\prime (r,\theta,\prime) \sin \theta d\theta d\phi \ .
\end{equation}
Horizontal lines in the top panels of Fig.~\ref{figadvection} mark the 
pressure level $p^\star$ at radius $r^\star$ where the contrast of $s^\star$ has decreased by 
$50$\%. This is roughly the depth where the term $-u_r \strat$ starts to 
balance Newtonian cooling. The decreasing Newtonian cooling and increasing 
balance with $-u_r \strat$ prevents an effective heating at larger depths and thus
a decrease in the term $-s^\prime/p^2\tau_0$ 
(green profiles in the top panels of Fig.~\ref{figadvection}). At $p=p^\star$ also the horizontal advection is irrelevant and the main balance is between the Newtonian Cooling and the radial thermal advection term.

Other panels in Fig.~\ref{figadvection} show maps of $s^\prime$ (second row) 
and the remaining terms in the entropy equation at pressure level $p^\star$. 
They illustrate the essential role of the term $-u_r \strat$ in determining 
the direction of the hotspot shift. The downstreams along the background entropy 
gradient provide the decisive heating that shifts the hotspot, 
eastward when the downstream is located eastward of the SSP and westward when
the downstream is located westward of the SSP. 
As explained above, the two scenarios go along with clockwise 
rotating northern circulation cells (Fig.~\ref{instab3} a) 
or anticlockwise northern cells (Fig.~\ref{instab3} b) respectively. 

 \begin{figure*}
     \centering
     \includegraphics[width=0.95\textwidth]{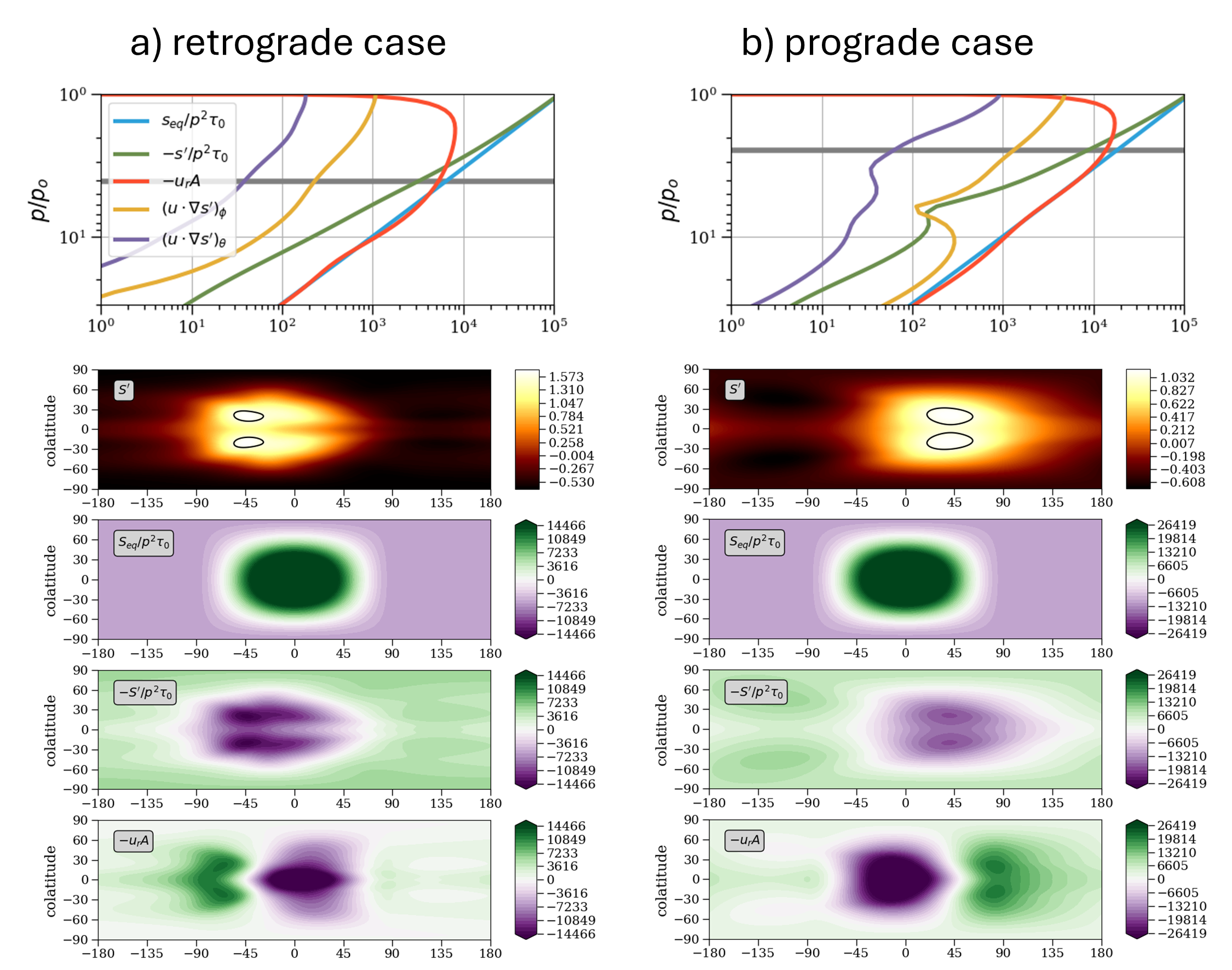}
     \caption{Radial profiles of the heating terms (top) for the retrograde (left) and prograde (right) hotspot shift. Following panels show maps of the entropy pattern (second row) 
     and the relevant terms of the entropy equation at the radius where the entropy contrast has been reduced to 50\%, i.e.~$r/r_o=0.971$ (left) and $0.982$ (right) . The maps show local heating in green and local cooling in purple. Parameters: a) $Ra=3 \cdot 10^7$, $\strat=10^3$ and b) $Ra=3 \cdot 10^8$, $\strat=10^3$. }
     \label{figadvection}
\end{figure*}

The entropy structure of our models can be related to photometric observations. 
Observations of IR phase curves yield information about the longitudinal offset of the brightest point and the relative day-to-nightside brightness ratio \citep{Parmentier2018, Bell2021}. 
The latter reflects the efficiency to equilibrate the entropy anomaly imposed by the irradiation. Observed is the total brightness, i.e.~the cumulative IR emission from several contributing layers weighted by the opacity of the IR wavelength. Observations on different wave lengths probe different depths.

IR emissions and the subsequent phase curves could, in principle, be modeled with 
radiative transfer codes. However, this is quite involved, as it requires a dedicated atmospheric $p$-$T$-model and detailed knowledge of the opacities, chemical abundances, the presence of haze or clouds and knowledge of the telescope specifications (wavelength bands etc.). \WD{This is beyond of the scope of this paper.}

Here we use a simpler approach and study the latitudinally averaged entropy anomaly $s^\star$ along depths in the equatorial plane: 
This is plotted in Fig.~\ref{fighotspiral} in the $p$-$\phi$-plane for 15 models at 
$\Ek=10^{-4}$ and  five Rayleigh numbers from $10^7$ to $10^9$ (horizontal direction) 
and $\strat$ values of $10^3$ (top), $10^2$ (middle), and $10$ (bottom).

Close to the top irradiation dominates and 
the entropy simply reflects our imposed equilibrium entropy pattern. 
Observationally, this would correspond to zero phase shift and maximum day-to-nightside brightness variation across all wavelengths.
The penetration depth of this pattern decreases when increasing either $\strat$ or $\Ra$. 
Both boost the relative important of $u_r \strat$ and leads to a 
more effective balance with Newtonian cooling. Increasing $\Ra$ simply yields
faster flows. Increasing $\strat$, on the other hand, yields slower flows, but since this 
is overcompensated by the increase in $\strat$ the product $u_r\strat$ still grows. 

Fig.~\ref{fighotspiral} also suggests that the hotspot shift changes from westward to eastward
when increasing the Rayleigh number. The dependence on $\strat$ is less clear.
The shift increases with depth, which means that observations should show a  
hotspot offset that increases with wavelength since larger wavelengths probe deeper. 

Fig.~\ref{fighotspiral2} collects pressure $p^\star$, the longitude of the equatorial downflow (minimal $u_r$) at $p^\star$, and the hotspot shift (longitude of the maximum in $s^\star$ at $p^\star$) for 39 cases that are in the super-laminar regime covering various stratifications (colors) and all studied Ekman numbers (symbol type). 
Panel a) shows that the penetration of the entropy disturbance decreases with growing 
$N/\Omega=\Ek \sqrt{\strat\Ra/\Pra}$ from rather deep reaching with $p^\star >10$ when $N/\Omega$ is between 10 and 30 to rather shallow with $p^\star < 2$ at $N/\Omega<300$.
Panel b) demonstrates that there are two longitude clusters for the strongest equatorial downflow around $-90^\circ$ and around $90^\circ$. Hence the radial inflows tend to be anchored at either of the day-night-terminators. The dependence on $N/\Omega$ 
is not obvious, but we find only western longitudes for $N/\Omega<3$ 
and only eastern longitudes for $H/\Omega>200$. 
Finally, panel c) confirms that the location of the $u_r$ minimum indeed 
determines the direction of the hotspot shift to a large degree, 
but not so much its absolute value. The hot spot shift is found by taking the longitude at which  $s^\star(p^\star)$ reaches its maximum.
Indeed, the majority of western downflow longitudes yields western hostspot shifts 
and actually all eastern downflow longitudes yield eastern hotspot shifts. 
}

\begin{figure*}
     \centering
     \includegraphics[width=\textwidth]{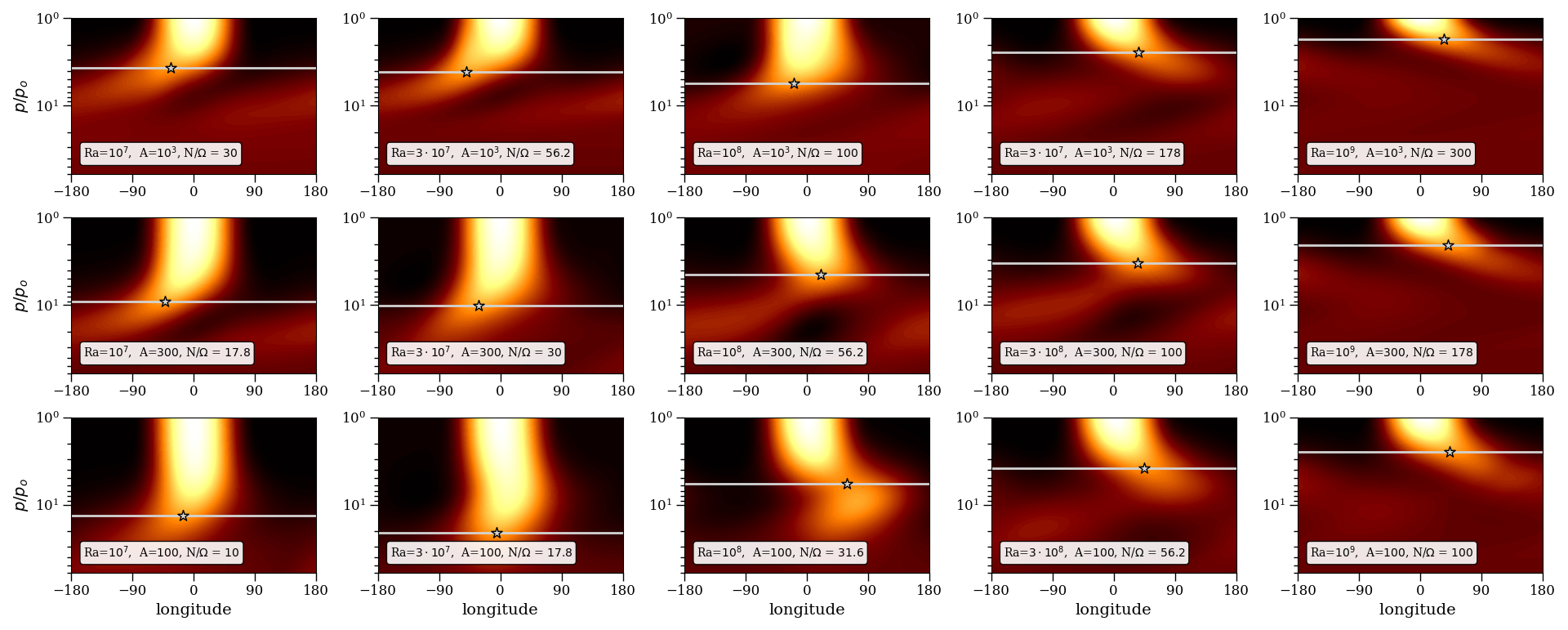}
     \caption{Latitudinally averaged entropy anomaly as a function of pressure and longitude for 15 different models. The pressure where the entropy contrast is dropped to 50\% is marked by the horizontal line. The position of the hotspot at this depth is indicated by the star. The pressure $p/p_o$ ranges from 1 to 60.}
     \label{fighotspiral}
\end{figure*}

\begin{figure*}
     \centering
     \includegraphics[width=\textwidth]{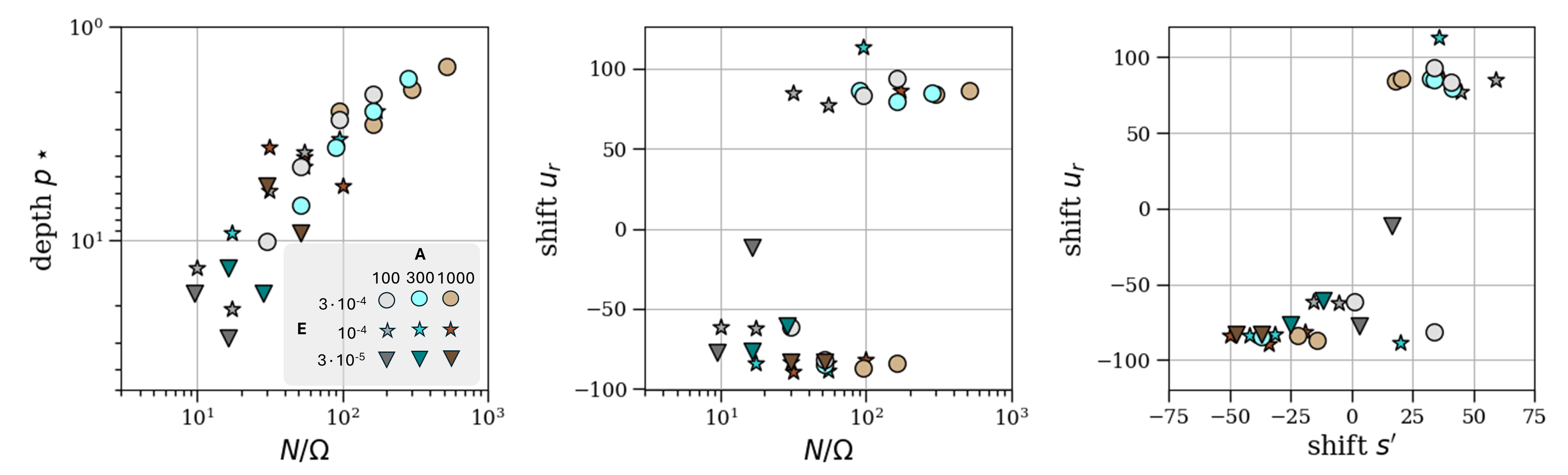}
     \caption{\WD{Depth of the heating anomaly (defined as the level where the entropy contrast has decreased to 50\%) as a function of $N/\Omega$, the shift of the maximum inward radial flow at the equator (shift $u_r$) as a function of $N/\Omega$, and the shift of $u_r$ as function of the shift of entropy maximum.} }
     \label{fighotspiral2}
\end{figure*}

\section{Discussion}
Hot Jupiters are giant planets that orbit their host stars in close proximity with synchronous rotation. The permanent dayside can reach temperatures of thousands of Kelvin, while the nightside radiates to space and remains much cooler. 
Thermal winds, driven by irradiation gradients, seek to equilibrate this thermal anomaly. At the depth where infrared emissions originate from, winds can smooth and shift the thermal anomaly. 

Observations allow us to deduce the brightness variations caused by infrared radiation during the planets orbit. 
Such infrared phase curves have revealed two unexpected findings. The first finding is that the hotspot 
is often shifted with respect to the substellar point. Prograde and retrograde shifts are observed, which can reach significant values.
Another key observation is the differences between day- and nightside emissions.
The difference is often significantly reduced 
compared to the purely radiative value, 
but the degree varies with
no apparent dependence on planetary or
stellar parameters. 
Both observations are key to
understanding the impact of the 
atmosphere dynamics.

So far, most numerical models have relied on generalized circulation models (GCMs) based on hydrostatic primitive equations, where radial flows arise from mass conservation rather than the full radial dynamics. Here, we propose an alternative approach that fully captures the three-dimensionality of the emerging flow. Rather than studying a specific Hot Jupiter, we focus on the general hydrodynamics of inhomogeneously irradiated gas planets and the emerging winds and thermal anomalies over a wide range of relevant parameters.

In our numerical simulations, we find very distinct flow regimes.
At moderate stratifications and larger Rayleigh numbers large scale Rossby waves appear and drift slowly 
westwards at a fraction of the rotation speed. 
Depending on the parameters, these waves can assume a meandering string of vortices very similar to what has been observed in Earth's atmosphere. Also possible are more sectorial patterns spanning from north to south. Inertial gravity waves of smaller scale and faster frequency are also found. Those are particularly effective in disturbing and destabilizing the radial flows into isotropic turbulence.

At the strong stratifications and large Rayleigh numbers that 
seem most applicable to HJs, the dynamics becomes super laminar and is dominated by strong azimuthal flows directed eastward at the equator 
and encircling the entire planet. 
These winds saturate at speeds of $\overline{u}_\phi = \Omega_P R_P$, i.e.~at Rossby numbers of order one, when the convective driving (Rayleigh number) is cranked up.
This simple rule implies that the wind speeds roughly equal the angular velocity of the tidally locked planet and predicts 
very fast velocities of several km/s for all Hot Jupiters, consistent
with the few available observations \citep{Snellen2010}. 

Such fast zonal flows are rather extreme. Even for the Sun, the differential rotation in terms of the Rossby number is about $Ro=0.3$. The differential rotation of the Solar System Gas Giants, where the zonal jets are driven by thermal convection, is an order of magnitude weaker at $Ro=0.012$ and $0.045$ for Jupiter and Saturn \citep{Aurnou2007} respectively. In numerical models of rotating convection, zonal winds reach comparable amplitudes of $Ro=0.05$ \citep{Wulff2022}. 
We attribute the large Rossby numbers found here to
the extreme one-sided irradiation. 

Due to the strong stable stratification, the radial flows are significantly slower, varying from 
some m/s to several tens of m/s, depending on the 
degree of stratification. This is consistent with independent assessments of vertical mixing by radial flows \citep{Komacek2019}.

Our analysis of the entropy pattern suggest that the longitudinal shift of the brightness maximum and the degree of day-to-nightside entropy contrast strongly depends on depth. 
While the shift can strongly increase with depth, the contrast
decreases at a rate that grows with the degree of 
stable stratification. 
The observation at a specific wave length 
will therefore depend on the depth at which the respective radiation predominately originates from.

 It also turns out that the radial heat advection can significantly impact the entropy pattern. Though the radial flows are generally slow, their impact is boosted by the fac\WD{t} that they act along the strong background stratification. Radial flows can cause both positive and negative brightness offsets with respect to the sub-stellar points. This depends on the  parameters, such as the orbital period or the surface gravity of the planet. We attribute these new findings to the use of a full 3D model in the anelastic approximation rather than the typical GCM approach of previous studies. \WD{However, the importance of radial advection has also been emphasized in GCM models \citep{Sainsbury-Martinez2023}. Here we find that it is the dominating effect in balancing the irradiation.}

Typical GCM simulations predict an eastward hotspot shift due to
the advection by the strong prograde zonal winds. 
However, there are several HJs for which westward hotspot advection was detected, namely WASP-33 b, WASP-12 b and Corot-2 b \citep{Dang2018, Bell2019, vonEssen2020, Bell2021}. This is typically attributed to electromagnetic effects, such as Lorentz forces disturbing the circulation \citep{Rogers2014, Hindle2021}. However, \citet{Dietrich2022} demonstrated that magnetic effects are only relevant when the equilibrium temperature exceeds 1500 K,
which has recently been confirmed by simulations using the 
approach presented here \citep{Boening2025}. 
We now demonstrate that retrograde or westward phase shifts are also possible without invoking electromagnetic effects, suggesting that westward hotspot offsets may be more common than previously expected.

\begin{acknowledgements}
The authors thank Vincent Böning and Paula Wulff for valuable discussions that significantly improved the manuscript. This work was supported by the Deutsche Forschungsgemeinschaft (DFG) in the framework of the priority program SPP 1992 ‘Exploring the Diversity of Extrasolar Planets’. The MAGIC code is available at an online repository (\url{https://github.com/magic-sph/magic}). 
    
\end{acknowledgements}

\bibliography{HotJup}{}
\bibliographystyle{aasjournalv7}

\end{document}